\documentclass[nofootinbib,amsmath,amssymb,aps,floatfix,notitlepage]{revtex4-1}
\pdfoutput = 1

\usepackage{subfigure}     
\newcommand{\be}{\begin{equation}}
\newcommand{\ee}{\end{equation}}

\usepackage{graphicx}

\usepackage[utf8]{inputenc}
\usepackage{mwe}
\usepackage{listings}
\usepackage{geometry}

\usepackage{amssymb}
\usepackage{amsmath}
\usepackage{epstopdf}
\usepackage{xcolor}
\usepackage{mathtools}
\usepackage{comment}

\def\hhref#1{\href{http://arxiv.org/abs/#1}{arXiv:#1}} 
\usepackage{hyperref}

\begin{document}  

\title{Bions and Instantons  in Triple-well and Multi-well Potentials}

\author{Gerald  V. Dunne}
\affiliation{Department of Physics, University of Connecticut, Storrs,  CT 06269-3046, USA}
\author{Tin Sulejmanpasic}
\affiliation{Department of Mathematical Sciences, Durham University, Durham, DH1 3LE, United Kingdom}
\author{Mithat \"Unsal}
\affiliation{Department of Physics, North Carolina State University, Raleigh, NC 27695, USA}

\begin{abstract} 
Quantum  systems with multiple degenerate classical harmonic minima exhibit new non-perturbative phenomena which are not present for the double-well and periodic  potentials. The simplest characteristic example of this family is  the triple-well  potential. Despite the fact that instantons are exact semiclassical solutions with finite and minimal action, they do not contribute to the energy spectrum at leading order in the semiclassical analysis. This is because the instanton fluctuation prefactor vanishes, which can be interpreted as the action becoming infinite quantum mechanically. Instead, the non-perturbative physics is governed by different types of {\it bion} configurations. A generalization to supersymmetric and quasi-exactly soluble models is also discussed.  An interesting  pattern of interference between topological and neutral bions, depending on the hidden topological angle, the discrete $\theta$ angle and the perturbative level number, leads to an intricate pattern of divergent/convergent expansions for low lying states, and provides criteria for  the exact solvability of some of the states. We confirm these semiclassical bion predictions using the Bender-Wu Mathematica package to study the structure of the associated perturbative expansions.  It also turns out that all the systems we study have a curious exact one-to-one relationship between the perturbative coefficients of the three wells, which we check using the BenderWu package. 
\\
\\
\centerline{\underline{\it  Dedicated to Roman Jackiw on the occasion of his 80th birthday}}
\end{abstract}

\maketitle

\section{Introduction}
\label{sec:intro}

Much of our physical intuition about instantons is derived from quantum mechanical instanton examples such as the symmetric double-well or periodic cosine potential, for which the instanton analysis is standard text-book material \cite{Coleman:1978ae,schulman,Vainshtein:1981wh,Balitsky:1985in,Peskin:1995ev,itep,ZinnJustin:2002ru,Schafer:1996wv,Marino:2015yie,Nekrasov:2018pqq}. The subject of instantons is  one in which Roman Jackiw has made many important contributions, both original discoveries \cite{Jackiw:1976pf,Jackiw:1976fs,Jackiw:1977pu}, and also crystal clear pedagogical expositions that get to the heart of the physics \cite{Jackiw:1977yn,Jackiw:1979ur,Jackiw:1983nv}. We dedicate this paper to Roman, in recognition of his profound influence on the many facets of non-perturbative physics. In this paper we describe several new  and intriguing aspects of instantons,  critical points at infinity and non-trivial configurations that live on their thimbles, known as bion configurations, that  arise in multi-well potentials with degenerate minima. The  simplest example that captures the new non-trivial aspects  of such systems is the  triple-well potential. 
 
It is first useful to recall the  salient features of the symmetric double-well potential,  both  the bosonic system and  its supersymmetric (SUSY) extension,   
to exhibit the sharp contrast with  triple-well and multi-well potentials. Interestingly,  many things that we learn  in textbooks concerning the degenerate double-well system are not general rules, but exceptions that arise in the limiting case of identical neighboring wells, i.e. neighboring wells which are related by a symmetry.  

The symmetric double well potential can be expressed as
\begin{eqnarray}
V=\frac{\omega^2}{2}x^2(x-1)^2
\label{eq:vsdw}
\end{eqnarray}
Familiar non-perturbative features of this symmetric double-well system include:
\begin{enumerate}
\item
In the semiclassical limit of deep wells, the low-lying states are split into doublets with exponentially small energy splitting, $\Delta E\sim e^{-S_I}$, associated non-perturbatively with the existence of instanton solutions that tunnel between the two minima. These instanton configurations have action $S_I$.
\item
Perturbation theory in each well is asymptotic, with expansion coefficients that are factorially divergent and non-alternating in sign: $c_n\sim \frac{n!}{(2S_I)^n}$. This behavior is associated with the existence of non-perturbative neutral-bion configurations, correlated instanton/anti-instanton configurations, which live on the thimble of critical points at infinity\cite{Behtash:2018voa}.

\item
In the SUSY extension of this model, the ground state energy vanishes to all orders perturbatively, but SUSY is broken non-perturbatively: $E_0^{NP}\sim - e^{-2S_I+i \pi }$. This behavior is associated with a non-perturbative neutral bion solution to the second order equations of motion associated with the classical plus quantum potential, $\frac{1}{2} (W')^2  \pm 
\frac{\hbar }{2} W'' $, 
having action $2S_I $ and hidden topological angle $ \theta_{\rm HTA} = \pi$.  In the original instanton language, this can be viewed as a configuration on the  Lefschetz thimble of  an instanton/anti-instanton critical point at infinity. 
  The quasi-zero-mode  thimble integral lives in   the complex plane, which is responsible for the  $ \theta_{\rm HTA} = \pi$, and which is the semiclassical origin of the  positivity of the non-perturbative energy shift.
\end{enumerate}

Many of the characteristic features of multiple-well systems appear already for the triple-well potential. For comparison purposes we concentrate on the symmetric triple-well potential (see Figure \ref{fig:tw-levels}):
\begin{eqnarray}
V=\frac{\omega^2}{2}x^2(x^2-1)^2
\label{eq:vdstw}
\end{eqnarray}
The most interesting non-perturbative features of this symmetric triple-well system are:
\begin{enumerate}
\item
There are exact instanton solutions tunneling between neighboring minima, with finite action $S_I$, but they play no role for the leading non-perturbative effects! This is due to the vanishing of the prefactor coming from the instanton fluctuation determinant. 
\begin{figure}[htbp] 
 \vskip -1.5cm  \centering
   \includegraphics[width=0.5\textwidth, trim=0 5 0 0]{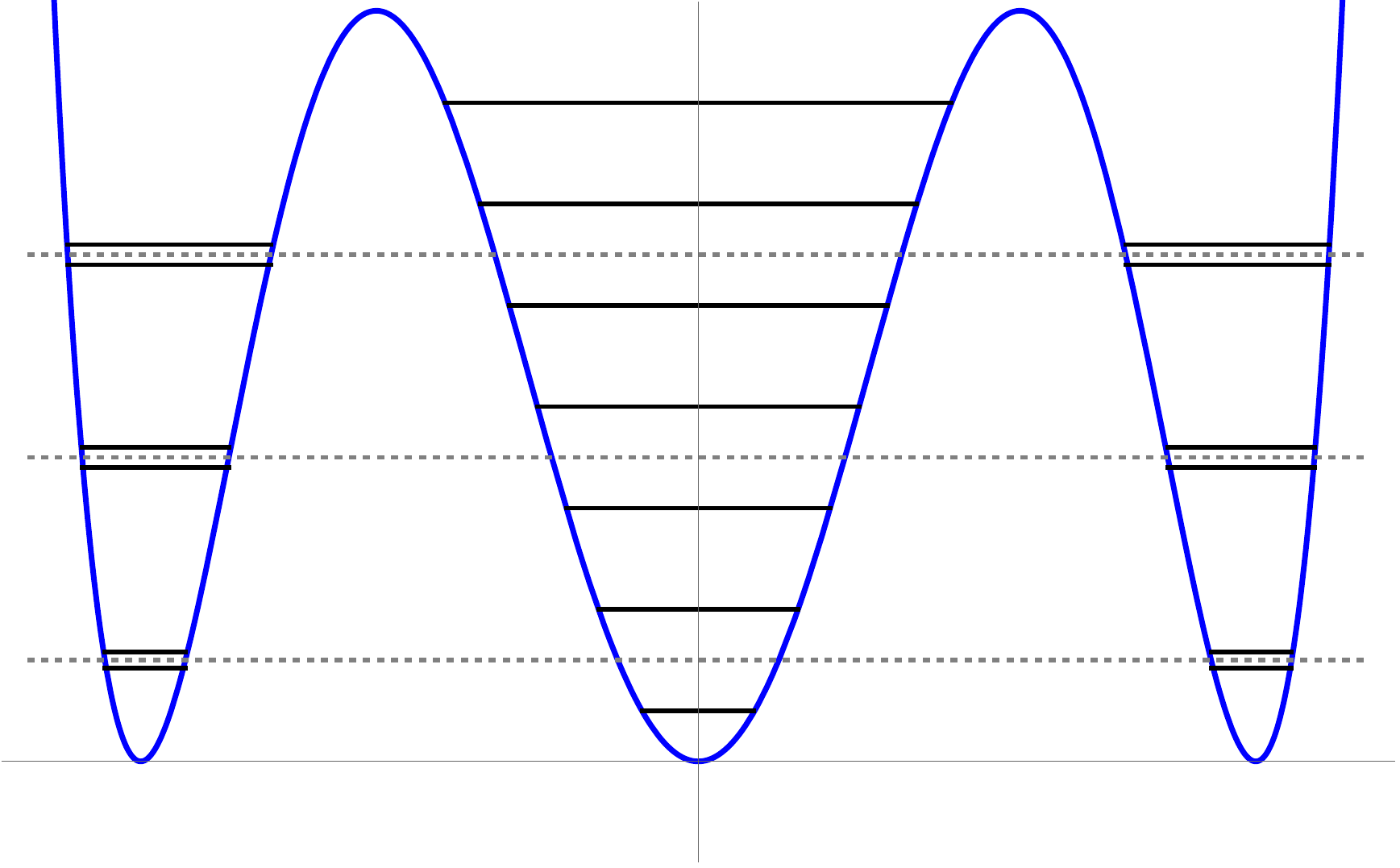}\includegraphics[width=.5\textwidth, viewport= 2.5cm 1.4cm 15cm 15cm,clip=true,height=5.6cm]{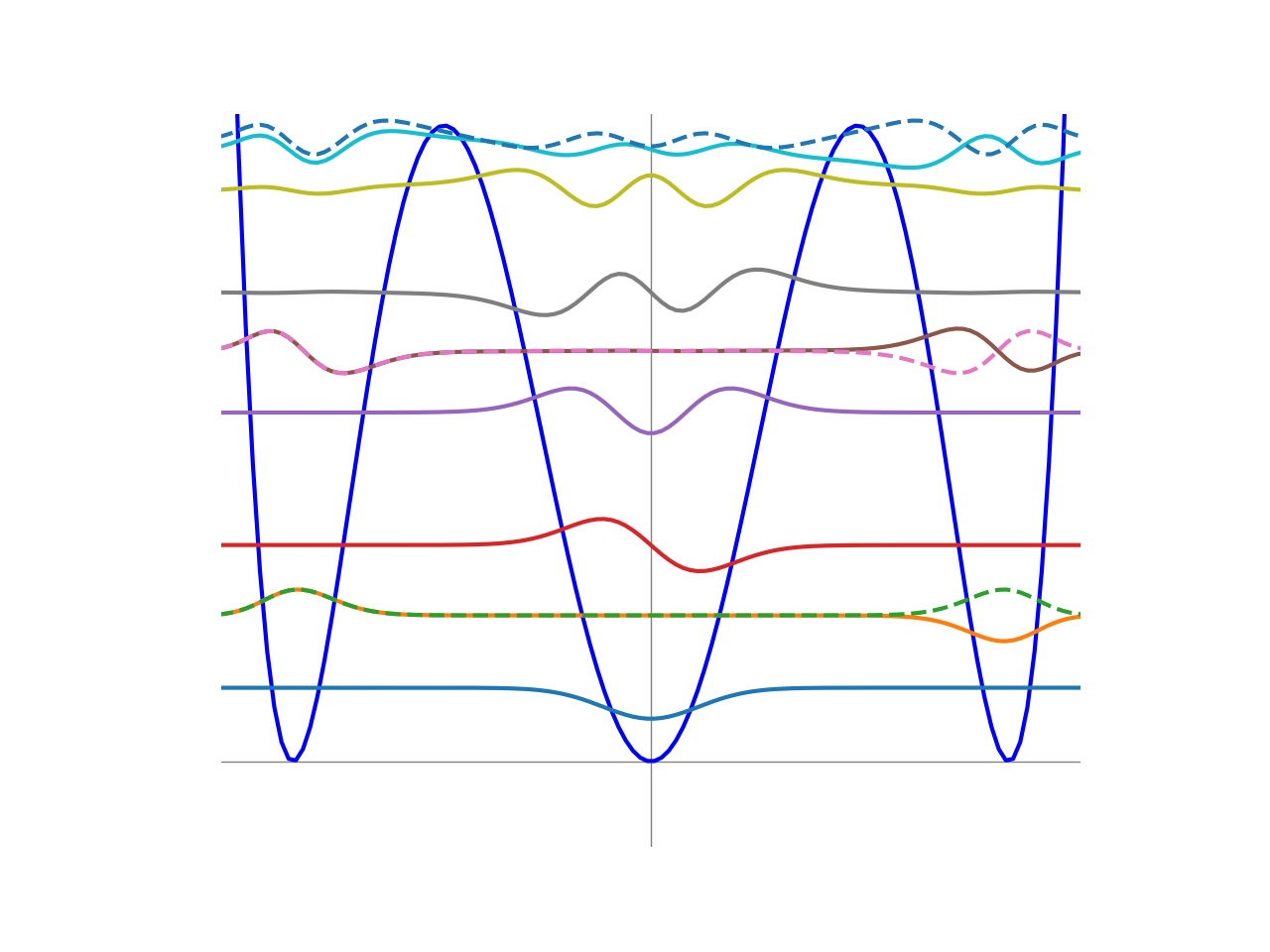} 
   \caption {Energy levels in the symmetric triple-well potential (\ref{eq:vdstw}). The outer wells have classical frequency $\omega_{\rm outer} =2\omega$, while the inner well has frequency $\omega_{\rm inner}=\omega$. The perturbative energy levels for states localized  in the outer wells are split into doublets by tunneling, while those for states localized in the inner well are not split into doublets. Note that the unperturbed harmonic levels for states localized in the inner and outer wells are interlaced in a systematic way. The right hand side shows the numerical solutions of the low-energy wavefunctions. Note that for nearly degenerate wave-functions we have represented the one with a higher energy with a dashed line. Notice also that for the lowest doublet the even, rather than the odd, eigenfunction has higher energy. }
   \label{fig:tw-levels}
\end{figure}
\item
The low-lying energy levels for states localized in the outer wells are exponentially split by a {\it two-instanton} effect rather than a one-instanton effect, $\Delta E_{\rm outer}\sim e^{-2S_I}$, while the energy levels for states localized in the inner well are not split at all, $\Delta E_{\rm inner}=0$. Rather, they are shifted up or down by the neutral bion contribution, 
$E_{\rm inner}^{\rm shift} \sim e^{-2S_I}$.   In the first doublet, the lower state has an {\it anti-symmetric} wavefunction, in contrast to the situation in the double-well potential where the lower state wavefunction is symmetric. In fact, the pattern of wavefunction symmetries is quite different, since the pattern must respect the oscillation theorem (increase of the number of nodes  with energy) and the parity symmetry properties of the wavefunctions. See Figure \ref{fig:tw-levels}.
\item
There are  two identical barriers, but two different types of wells: inner and outer. See Figure \ref{fig:tw-levels}. Nevertheless,  the perturbative expansions in the inner and outer wells are explicitly related: see Eq. (\ref{eq:pt-relation}) below.
\item
Perturbation theory in each kind of well is asymptotic, with expansion coefficients that are factorially divergent and non-alternating in sign: $c_n\sim \frac{n!}{(2S_I)^n}$. This behavior is associated with the existence of non-perturbative neutral-bion configurations  with zero topological charge but action equal to twice the instanton action. 
These can be interpreted as configurations on the thimble of an instanton/anti-instanton critical point at infinity. 

\item
In the SUSY extension of this triple-well model, the ground state energy vanishes to all orders perturbatively for both $(-1)^F$ even/odd sectors $H^{\pm}$.   In  the $H^{-}$ sector,  the ground state energy is $E_0^{NP}=0$, due to the fact that there is a normalizable zero-mode of $H^{-}$. Semiclassically this vanishing of $\Delta E_0^{NP}$ is due to cancellation between topological and neutral bion contributions;  while in the $H^{+}$ sector, $E_0^{NP}\sim  - e^{-2S_I +i\pi}$, due to an unpaired neutral bion. The hidden topological angle is crucial for both these results. 

\item
In the SUSY extension there is again a simple explicit relation between the perturbative expansion coefficients in the inner and outer wells. See Eq. (\ref{eq:susy-pt-relation}).

\item
There is a $\zeta$-deformed generalization\cite{Balitsky:1985in,Verbaarschot:1990ga,Verbaarschot:1990fa,Behtash:2015loa} of the SUSY triple-well system, discussed in Section \ref{sec:zeta}, in which the fermion number parameter is deformed from $\zeta=1$.
When $\zeta= \frac{2m+1}{3},  m=1, 2,  \ldots $ curious cancellations arise in the semiclassical analysis, and result in convergent perturbation theory for part of the spectrum.  These special $\zeta$ values correspond to the quasi-exactly solvable potentials.\cite{Turbiner:1987nw,Turbiner:1987kt,Turbiner:1994gi}
Namely the corresponding  $H^{-, \zeta}$ system has $2m$ lowest states for which  perturbation theory is convergent, while for higher states one finds the generic divergent perturbative expansion.  Of these $2m$ states, $m$ are exactly solvable by methods of \cite{Turbiner:1987nw,Turbiner:1987kt,Turbiner:1994gi}, with a convergent perturbative expansion which sums to the correct result. In this case the two types of bion contribution cancel against each other. The other $m$ states are not exactly solvable, and there is a non-vanishing combination of topological and neutral bion contributions.  For $H^{+, \zeta}$, there is an alternating pattern for the lowest $m$ states. $ \lceil \frac{m}{2} \rceil $ of these states have convergent perturbative expansions and non-perturbative contributions from neutral and topological bions, and $m- \lceil \frac{m}{2} \rceil $ have asymptotic divergent expansions. None of these states are exactly solvable non-perturbatively. 
 The interplay between the topological and neutral bions provides a path integral semi-classical explanation of certain puzzles about non-perturbative effects in quasi-exactly solvable systems, for which a finite number of energy levels can be found algebraically.
\end{enumerate}

These physical features of the triple-well potential (\ref{eq:vdstw}) are explained semiclassically in the following sections. They illustrate that the double-well potential (\ref{eq:vsdw}) is in fact quite special, in the sense that neighboring wells have the same frequency, and this is what makes instanton solutions physically relevant. In the case for which neighboring wells do not have the same frequency, such as the triple well potential (\ref{eq:vdstw}), despite the fact that the instanton action is finite, the instanton amplitude is zero due to the vanishing of the fluctuation prefactor.

\section{Instantons and Bions in the Triple Well System}
\label{sec:sdtw-bions}

\subsection{Vanishing of Instanton Amplitudes for Inequivalent  Degenerate Wells}
\label{sec:dstw-insts}

In this Section we show that given two consecutive degenerate harmonic wells with frequencies $\omega_1 \neq \omega_2$,  despite the fact that a classical finite action instanton solution exists for the Euclidean BPS equation, 
the amplitude for such an instanton is zero due to quantum mechanical effects. 
Therefore, the instantons in general do not contribute to the energy spectrum at leading order in semi-classics, even though they are exact solutions with minimal action.  
 Below, we describe this effect for  the symmetric triple-well potential, which captures the essence of the general case. 

The BPS equations for the symmetric triple-well potential (\ref{eq:vdstw}), $\dot{x}=\mp \omega \, x(x^2-1)$, have instanton and anti-instanton solutions:
\begin{eqnarray}
{\rm instantons:}\quad x_I^{(\pm)}(t)&=& \pm \frac{1}{\sqrt{1+e^{-2\omega (t-t_0)}}}\\
{\rm anti-instantons:}\quad x_{\bar{I}}^{(\pm)}(t) &=& \pm \frac{1}{\sqrt{1+e^{2\omega (t-t_0)}}}
\label{eq:dstw-bps}
\end{eqnarray}
\begin{figure}[htbp] 
   \centering
     \includegraphics[width=0.65\textwidth]{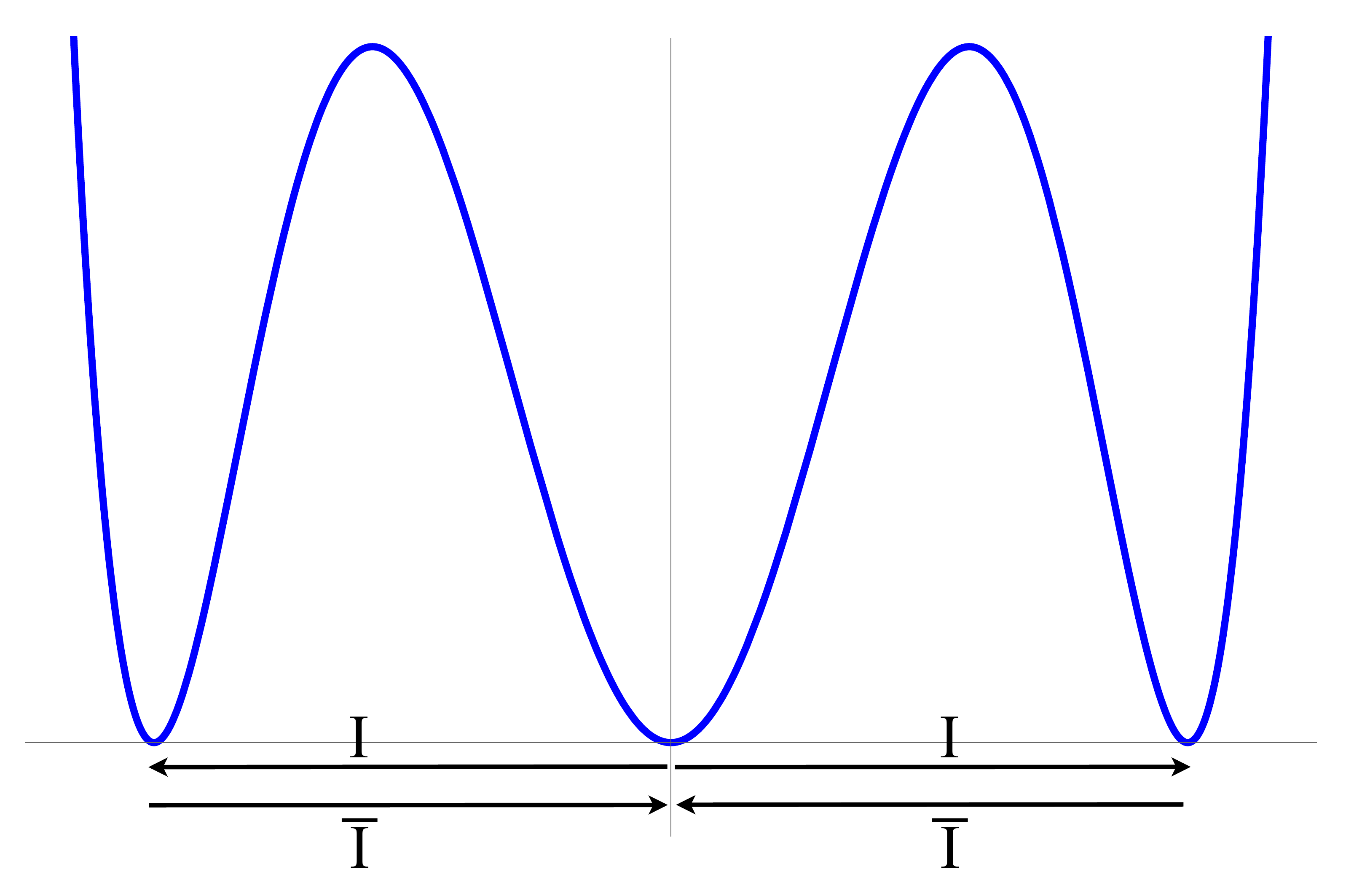}
   \caption{ Instantons and anti-instantons in the symmetric triple well  (\ref{eq:vdstw}).  
   Instantons tunnel from the inner minimum to the outer minima $(\pm)$,  and anti-instantons tunnel from the outer minima $(\pm)$ to the inner minimum. 
   } 
   \label{fig:inst}
\end{figure}
Here, the $(\pm)$ notation denotes the minimum (at $x=\pm 1$) to or from which the solution tunnels, and  $t_0$ denotes the zero-mode degree of freedom. The instanton solutions $x_I^{(\pm)}(t)$ tunnel from the inner vacuum at $x=0$ to the outer vacua at $x=\pm 1$, while the anti-instanton solutions $x_{\bar{I}}^{(\pm)}(t)$ tunnel from the outer vacua at $x=\pm 1$ to the inner vacuum at $x=0$. See Figure \ref{fig:inst}. The instanton and anti-instanton action  is 
\begin{eqnarray}
S_I=S_{\bar{I}}= \frac{\omega}{4}
\label{eq:si}
\end{eqnarray}
The quadratic fluctuation  operator in the background of an  instanton is $F=(-\partial_t^2+V_{\rm fluc}(t))$ where
\begin{eqnarray}
V_{\rm fluc}(t)&=& \left[V^{\prime\prime}(x)\right]_{x=x_I^{(\pm)}(t)}\nonumber\\
&=& \frac{\omega^2 \left(-10 \, e^{2\omega (t-t_0)}+4 \, e^{4 \omega (t-t_0)}+1\right)}{\left(e^{2 \omega (t-t_0)}+1\right)^2}.
\label{eq:vfluc}
\end{eqnarray}
\begin{figure}[htbp] 
   \centering
     \includegraphics[width=0.75\textwidth]{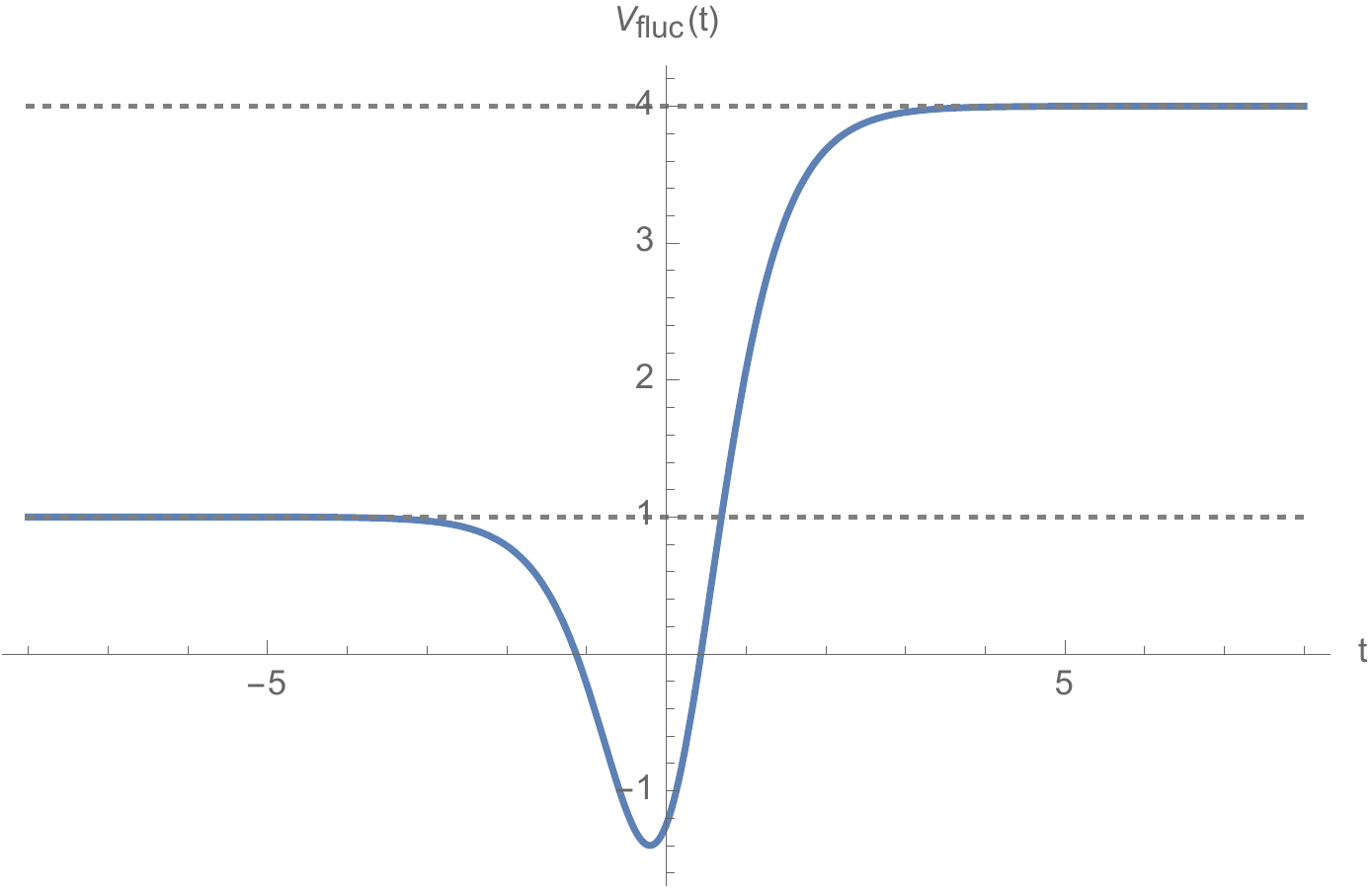}
   \caption{The instanton fluctuation operator potential   (\ref{eq:vfluc}) for the symmetric triple well.  
 The mis-match between the $t\to\pm\infty$ limits leads to a vanishing fluctuation determinant: see Eq. (\ref{eq:inst-det}). 
   } 
   \label{fig:inst-fluc}
\end{figure}
See Figure \ref{fig:inst-fluc}. (For an anti-instanton: $t\to -t$.)
The non-perturbative amplitude associated with an instanton or anti-instanton is  a standard textbook computation \cite{Coleman:1978ae,schulman,Vainshtein:1981wh,Balitsky:1985in,Peskin:1995ev,itep,ZinnJustin:2002ru,Schafer:1996wv}:
\begin{eqnarray}
\left[ I\right] \sim   J _{\tau_0}  \left[ \frac{\det^\prime F}{\det F_0} \right]^{-1/2} e^{-S_I}
\label{eq:inst-amp}
\end{eqnarray}
Here,  prime indicates that the zero mode is omitted from the determinant, as it must be  integrated over exactly. The Jacobian factor is  given by $J _{t_0}  = \sqrt{ \frac{S_I}{2\pi} }$, and  ${\det F_0}$ is the normalization by the free fluctuation operator, which if one is interested in the ground state should have the frequency of the middle well.  The determinant factor of this instanton amplitude can be expressed in terms of asymptotic values of the instanton and anti-instanton solutions \cite{Coleman:1978ae,ZinnJustin:2002ru,Dunne:2007rt}:
\begin{eqnarray}
\left[ \frac{\det^\prime F}{\det F_0} \right]^{-1/2}  \sim   \lim_{\beta \to\infty}e^{- \frac{\beta}{2} \times \frac{1}{2}(\omega_{1}-\omega_{0})} =0
\label{eq:inst-det}
\end{eqnarray}
Here $\omega_0=\omega$ is the frequency of the inner well, and $\omega_1=2\omega$ is the frequency of the outer wells, and $\beta$ is the regulated length of the Euclidean time direction. The determinant factor in (\ref{eq:inst-det}) vanishes, due to the mis-match of the frequencies in the inner and outer wells. Therefore
\begin{eqnarray}
\left[ I\right] = 0 = \left[ \bar{I}\right] 
\label{eq:inst-amp-zero}
\end{eqnarray}
This means that the instanton solutions, despite being the non-perturbative objects with the lowest action, do not contribute to the energy spectrum of the ground state and low-lying states. This is also clear from looking at the numerical wavefunctions shown in Fig.~\ref{fig:tw-levels}, where the states are either localized in the outer wells or the inner wells, but not in both.  Generalizing this argument, we see that this is actually true whenever classical degeneracy is not accidental, i.e. not related by a symmetry. 

\subsection{Perturbation Theory in the Triple Well System}
\label{sec:dstw-pt}

There are two different wells in the triple well potential (\ref{eq:vdstw}), so we might expect that perturbation theory is different in the inner and outer wells. Indeed, the perturbative expansions for low-lying levels {\it look} different. We can study the various perturbative expansions using the BenderWu Mathematica package \cite{Sulejmanpasic:2016fwr}. Comparing the perturbative expansion coefficients for the first few levels in the inner and outer wells suggests no particular relation between them beyond the obvious relation for the first term, the unperturbed energy:

\noindent \underline{inner well perturbative expansion coefficients:}
\begin{eqnarray}
 \nu_{\rm inner} =0  &:&
\left\{\frac{1}{2},-\frac{3}{4},-\frac{27}{16},-\frac{153}{16},-\frac{20385}{256},
-\frac{27027}{32},
\dots
   \right\}\\
 \nu_{\rm inner} =1&:&
  \left\{\frac{3}{2},-\frac{15}{4},-\frac{225}{16},-\frac{2025}{16},-\frac{411075}{256},-\frac{799875}{32},
  \dots
   \right\} \\
\nu_{\rm inner} =2&:&
   \left\{\frac{5}{2},-\frac{39}{4},-\frac{855}{16},-\frac{10809}{16},-\frac{3009285}{256},-\frac{7884891}{32},
    \dots
   \right\} 
\label{eq:inner-data}
\end{eqnarray}

\noindent \underline{outer well perturbative expansion coefficients:}
\begin{eqnarray}
\nu_{\rm outer}=0 &:&
\left\{1,-\frac{15}{8},-\frac{45}{8},-\frac{5265}{128},-\frac{6885}{16},-\frac{5735205}{1024},
\dots
\right\}\\
\nu_{\rm outer}=1 &:&
 \left\{3,-\frac{111}{8},-\frac{711}{8},-\frac{165969}{128},-\frac{412695}{16},-\frac{628455429}{1024},
 \dots
 \right\} \\
\nu_{\rm outer}=2 &:&
  \left\{5,-\frac{303}{8},-\frac{3105}{8},-\frac{1132497}{128},-\frac{4310145}{16},-\frac{9871632837}{1024}, \dots
   \right\}
\label{eq:outer-data}
\end{eqnarray}
To recognize the relation between the perturbative expansions in the inner and outer wells, we write the expansion coefficients in terms of the perturbative level number $\nu$ for the respective well.
The first five orders of the perturbative expansions for the inner and outer wells, as a function of the perturbative level number $\nu$ in the respective well, are:
\begin{eqnarray}
{\rm inner}&:& \left\{\nu+\frac{1}{2},-\frac{3 \nu^2}{2}-\frac{3 \nu}{2}-\frac{3}{4},-3 \nu^3-\frac{9 \nu^2}{2}-\frac{39 \nu}{8}-\frac{27}{16}, 
\right.\nonumber\\
&&\left.
-\frac{105
   \nu^4}{8}-\frac{105 \nu^3}{4}-\frac{363 \nu^2}{8}-\frac{129 \nu}{4}-\frac{153}{16},
    \right.\nonumber\\
&&\left.
   -\frac{603 \nu^5}{8}-\frac{3015 \nu^4}{16}-\frac{3645
   \nu^3}{8}-495 \nu^2-\frac{39897 \nu}{128}-\frac{20385}{256}, \dots
   \right\}
    \label{eq:Ncoeff-inner}
\end{eqnarray}
\begin{eqnarray}
{\rm outer}&:&
\left\{2 \nu+1,-6 \nu^2-6 \nu-\frac{15}{8},-24 \nu^3-36 \nu^2-\frac{93 \nu}{4}-\frac{45}{8},
\right.\nonumber\\
&&\left.
-210 \nu^4-420 \nu^3-\frac{1671 \nu^2}{4}-\frac{831
   \nu}{4}-\frac{5265}{128},
      \right.\nonumber\\
&&\left.
   -2412 \nu^5-6030 \nu^4-\frac{16335 \nu^3}{2}-\frac{24885 \nu^2}{4}-\frac{20259 \nu}{8}-\frac{6885}{16}, \dots
   \right\}
   \label{eq:Ncoeff-outer}
\end{eqnarray}

These inner-well and outer-well expansion coefficients do not immediately look like they are related, but when expressed as functions of the action parameter, $B\equiv \nu+\frac{1}{2}$, we find:
\begin{eqnarray}
{\rm inner}&:&
 \left\{B,-\frac{3 B^2}{2}-\frac{3}{8},-3 B^3-\frac{21 B}{8},-\frac{105 B^4}{8}-\frac{411 B^2}{16}-\frac{297}{128}, \right.\nonumber\\
&&\left.
-\frac{603
   B^5}{8}-\frac{4275 B^3}{16}-\frac{351 B}{4}, \dots
   \right\}
  \label{eq:Bcoeff-inner} \\
   {\rm outer}&:&
\left\{2 B,-6 B^2-\frac{3}{8},-24 B^3-\frac{21 B}{4},-210 B^4-\frac{411 B^2}{4}-\frac{297}{128},
\right.
\nonumber\\
&& \left. -2412 B^5-\frac{4275 B^3}{2}-\frac{351
   B}{2}, \dots
   \right\}
  \label{eq:Bcoeff-outer}
   \end{eqnarray}
We recognize that the perturbative expansions are related by a simple explicit map:
\begin{eqnarray}
E_{\rm pert}^{\rm outer}(B, \hbar)=E_{\rm pert}^{\rm inner}(2B, \hbar)
\label{eq:pt-relation}
\end{eqnarray}
We have verified this result to very high orders using the BenderWu package \cite{Sulejmanpasic:2016fwr} for computing perturbative expansions.

The origin of relation (\ref{eq:pt-relation}) is not immediately obvious using direct Rayleigh-Schr\"odinger perturbation theory, but it can be understood straightforwardly using the relation to exact WKB. In this approach, ordinary perturbation theory can be generated by the following procedure \cite{ZinnJustin:2004ib,Basar:2017hpr}. First, compute the formal series for the all orders WKB action
\begin{eqnarray}
a(E, \hbar) =\sum_{n=0}^\infty \hbar^{2n} a_{2n}(E)
\label{eq:ao-wkb}
\end{eqnarray}
where $a_{2n}(E)$ are the WKB actions \cite{dunham,bender-book}:
\begin{eqnarray}
a_0(E)=\sqrt{2} \oint_{\rm tp} dx\, \sqrt{E-V} \quad; \quad a_2(E)=-\frac{\sqrt{2}}{2^6}\oint_{\rm tp} dx\, \frac{(V')^2}{(E-V)^{5/2}}\quad; \quad \dots 
\label{eq:wkbas}
\end{eqnarray}
In the triple-well system, all the $a_{2n}(E)$ are expressed in terms of simple hypergeometric functions.\cite{Basar:2017hpr}
Next, impose the all-orders Bohr-Sommerfeld quantization condition
\begin{eqnarray}
a(E, \hbar)=2\pi\hbar \left(\nu+\frac{1}{2}\right) 
\qquad, \quad \nu=0, 1, 2, \dots
\label{eq:bs}
\end{eqnarray}
Finally, expand each $a_{2n}(E)$ at small $E$ and then invert (\ref{eq:bs}) to express the energy as a function of $\nu$ and $\hbar$. This inversion produces an expression of the form $E=E(\nu, \hbar)$, rather than (\ref{eq:bs}) which is of the form $\nu=\nu(E, \hbar)$. Furthermore, the expression $E=E(\nu, \hbar)$ coincides with standard Rayleigh-Schr\"odinger perturbation theory expanded about the $\nu^{\rm th}$ harmonic unperturbed state. This perturbative procedure can be implemented for actions with turning points for the inner well or outer wells, and the coefficients of the $\hbar$ expansion are polynomials in the respective $\nu$ label, as listed in Eq. (\ref{eq:Ncoeff-inner}-\ref{eq:Ncoeff-outer}). 

The triple-well potential in (\ref{eq:vdstw}) has the interesting geometric property \cite{Basar:2017hpr} that at each order of the WKB expansion, the actions are {\it equal} in the inner and outer wells, up to a simple factor of 2:
\begin{eqnarray}
a_{2n}^{\rm inner}(E)=\frac{1}{2} a_{2n}^{\rm outer}(E)\qquad, \quad n=0, 1, 2, \dots
\end{eqnarray}
For example, at the classical level it is clear that the frequencies and actions in the inner and outer wells differ only by  factors of 2. Remarkably, this behavior persists to all orders for the triple-well system \cite{Basar:2017hpr}. This means that in this all-orders Bohr-Sommerfeld approach to perturbation theory, the symmetry under $B\to2B$ in (\ref{eq:pt-relation}) follows immediately from (\ref{eq:bs}), which is equivalent to perturbation theory.  
 
It is well known that important connections between perturbative and non-perturbative physics are encoded in the large-order behavior of perturbation theory \cite{Bender:1969si,Bender:1971gu,LeGuillou:1990nq,Brezin:1977gk}. For the triple-well potential (\ref{eq:vdstw}) the high orders of the perturbative expansions can be studied efficiently using the BenderWu Mathematica package \cite{Sulejmanpasic:2016fwr}. It is simple to generate many hundreds of terms in these expansions, which permits high-precision analysis  of the large-order growth. The leading growth for the perturbative expansion coefficients  in  (\ref{eq:Ncoeff-inner}-\ref{eq:Ncoeff-outer})  is as follows:
\begin{eqnarray}
\text{inner :} \qquad c_n^{({\rm level}\, \nu),\, {\rm inner}}&\sim&\beta_\nu\, 
 \frac{2^{n+\frac{3\nu}{2}-\frac{1}{4}}\,\Gamma\left(n+\frac{3\nu}{2}-\frac{1}{4}\right)}{\pi\, \Gamma\left(\frac{3\nu}{2}+\frac{3}{4}\right)}\qquad , \quad n\to \infty 
 \label{eq:asym-inner}\\
\text{outer :} \qquad c_n^{({\rm level}\,\nu),\, {\rm outer}}&\sim&
\gamma_\nu\, 
 \frac{2^{n+3\nu+\frac{1}{2}}\,\Gamma\left(n+3\nu+\frac{1}{2}\right)}{\pi\, \Gamma\left(3\nu+\frac{3}{2}\right)}\qquad , \quad n\to \infty
  \label{eq:asym-outer}
\end{eqnarray}
up to some $n$-independent rational normalization factors.
Note that these asymptotic behaviors are consistent with the exact relation (\ref{eq:pt-relation}), and the expression (\ref{eq:asym-inner}) is consistent with an expression in \cite{Brezin:1977gk}.

For our purposes here, the most important facts about large-order perturbation theory for the triple-well system are:
\begin{itemize}
\item the coefficients grow factorially in magnitude;
\item  the coefficients do not alternate in sign; 
\item the factor $2^n$ in (\ref{eq:asym-inner})-(\ref{eq:asym-outer}) corresponds to $1/(2S_I)^n$ in our normalization, the $n^{\rm th}$ power of the inverse of {\it twice} the instanton action \eqref{eq:si}.
\end{itemize}
 These facts imply that naive Borel summation of these formal perturbative series produces ambiguous imaginary non-perturbative terms with exponential factor $e^{-2S_I}$. These are cancelled by contributions from instanton/anti-instanton interaction effects \cite{Bogomolny:1980ur,ZinnJustin:1981dx,Behtash:2015zha,Behtash:2015loa,Behtash:2018voa,Fujimori:2017oab}, which we refer to as quantum bions, and which are analyzed in the next subsection. This cancellation mechanism is one of the clearest examples of the application of resurgent trans-series in physics, where a  trans-series combines both perturbative and non-perturbative contributions in such a way that naive imaginary terms are cancelled in the full trans-series, producing real and unambiguous physical results \cite{Marino:2012zq,Aniceto:2013fka}.

\subsection{Critical Points at Infinity, Lefschetz Thimbles and Bions in the Triple Well System}
\label{sec:dstw-bions}

In the two-instanton sector of a generalized instanton gas analysis,  we must carefully consider the critical points at infinity, and their Lefschetz  thimble contribution.\cite{Behtash:2018voa} The main results of Ref. \cite{Behtash:2018voa}  
for the double well-potential and its $\hbar$-tilting are the following: 
\begin{itemize}
\item{An instanton and anti-instanton is a critical point at infinite separation.  
Starting with  the theory  compactified on a circle with size $\beta$,  the classical + quantum interaction  between the two is of the form 
$V_{\rm eff} (\tau)=  \left( - A ( e^{-\omega\, \tau } +e^{-\omega (\beta- \tau) })   + \hbar \omega \tau \right)$. 
 The critical point is determined by the classical action, and located at  $\omega\, \tau^*=  \frac{\beta}{2}$. 
 As $\beta \rightarrow \infty$, the classical interaction between the two instantons dies off and the configuration becomes a genuine saddle point. }
 
 \item{ The thimble associated with the critical $\omega\, \tau^*$ is given by 
 (for $\hbar \rightarrow  e^{i \theta}\,\hbar$)
\begin{align}
&\Gamma_{\rm QZM}^{\theta=0^{+}}  = \gamma_1^+  + \gamma_2^+ +  \gamma_3^+
\end{align}
where the segments are $\gamma_1^+ =  (-\infty + i\pi,  \frac{\beta}{2} + i\pi )$, 
  $\gamma_2^+ = [ \frac{\beta}{2} + i\pi,  \frac{\beta}{2} - i\pi ]$,  
  $\gamma_3^+ = ( \frac{\beta}{2} - i\pi,\infty - i\pi ) \, . $  
In the $\beta \rightarrow \infty$ limit, the integration is equal to the contribution of the $\gamma_1^+$ segment. }

\item{These  critical points at infinity  are non-Gaussian. Therefore,  to 
 reproduce the correct NP-contributions at second order in semi-classics, the quasi-zero-mode (QZM) integrals need to be done exactly, not in the Gaussian approximation.}

\item{The configurations that dominate the integration over $\Gamma_{\rm QZM}^{\theta=0^{+}} $ are the   neutral and topological  bion configurations. } 
\end{itemize}

In a semi-classical treatment of the quantum mechanical path integral, we must take  into account the physical effect of  all  saddles and their thimbles, including the critical points at infinity,   and their thimbles:
see e.g.  \cite{Richard:1981gn,Lapedes:1981tz,Behtash:2015zha,Behtash:2015loa,Behtash:2018voa,Fujimori:2017oab,Dunne:2016nmc}. In what follows we analyze the triple well system using the notions of topological bions and neutral bions.

There are two different kinds of correlated instanton configurations:  topological\footnote{In some of our previous work these were called the ``real bions''. We have renamed them ``topological bions'' here because they need not be real saddles of the classical equations of motion. Their most important distinguishing feature is that they have a nonzero topological charge, in contrast to neutral bions whose topological charge vanishes.}
bions and neutral bions. We begin by discussing neutral bions of the classical triple-well system. These can be thought of as the solutions of the inverted potential which roll down from one of the peak of the inverted potential, to an adjacent peak of the inverted potential, and then roll back. It is clear that we can find a solution which performs this motion any number of times in time $\beta$. We will refer to an object which performs this motion once (down the hill, up the hill and back) as a neutral bion. The term neutral  indicates that there is no topological charge associated with this object.

The topological bion is more interesting. Intuitively such an object corresponds to a tunneling event between the far left well and the far right well. Indeed looking at Fig.~\ref{fig:tw-levels}, such an object must exist on physical grounds in order to explain the energy splitting between the states localized on the left and on the right. The name topological  emphasizes the topological stability of the configuration which has its two ends in two classical minima related by a symmetry. However the classical potential clearly does not have real finite action solutions of this kind\footnote{Thinking of the inverted potential of the triple-well, the only classical solutions which go from one outer maximum to another would be flying off to infinity, and would hence have infinite action.}.
Candidate solutions of the topological bion can be found in various ways as limits approaching this functional separatrix. One possibility is to make a small shift of the minimum of the middle well, in which case a solution appears. Another possibility is to take a limit of the complex solutions at energy just above the top of the inverted potential. 
Another approach, which we adopt here, is to employ minimization of the quantum action, which yields physically consistent results.\cite{Behtash:2015zha,Behtash:2015loa,Fujimori:2017oab}. These quantum saddles correspond to the tails of the Lefshitz thimbles of the appropriate saddles minimizing the classical action, in the spirit of \cite{Behtash:2018voa}. The full details of the connection between the aforementioned limiting procedures is interesting and deserves further future study.
However, for the purposes of this work we consider both the neutral bion and the topological bion to be solutions of the equations of motion with the quantum corrected potential $\frac{1}{2} (W')^2 +  \frac{\hbar}{2} W''$,  as in Refs.\cite{Behtash:2015zha,Behtash:2015loa,Fujimori:2017oab}. 
The neutral bions are also well-defined objects on the thimble as in \cite{Behtash:2018voa}, and we expect that the same holds for the topological bions. We show that this combined approach  is self-consistent.  

The characteristic size  of bions in the Euclidean time direction is given by 
\begin{eqnarray}
t_*\sim \ln \frac{1}{\hbar}
\label{eq:tstar}
\end{eqnarray}
The topological and neutral bions have finite action, and their fugacities are given by
\begin{eqnarray}
[\mathcal{TB}]\propto e^{-2S_I}\qquad,\qquad [\mathcal{NB}]\propto e^{-2S_I \pm i\frac{\pi}{4}}
\label{eq:bion-actions}
\end{eqnarray}
for the topological bion $[\mathcal {TB}]$ and the neutral bion $[\mathcal{NB}]$ respectively. 
The complex phase, the hidden topological angle $\theta_{HTA}$, comes from the fact that these configurations are in fact complex configurations which saturate the thimble integral. The physical effects of the topological and neutral bions are very different.
\begin{figure}[htb]
\centering\includegraphics[scale=.6]{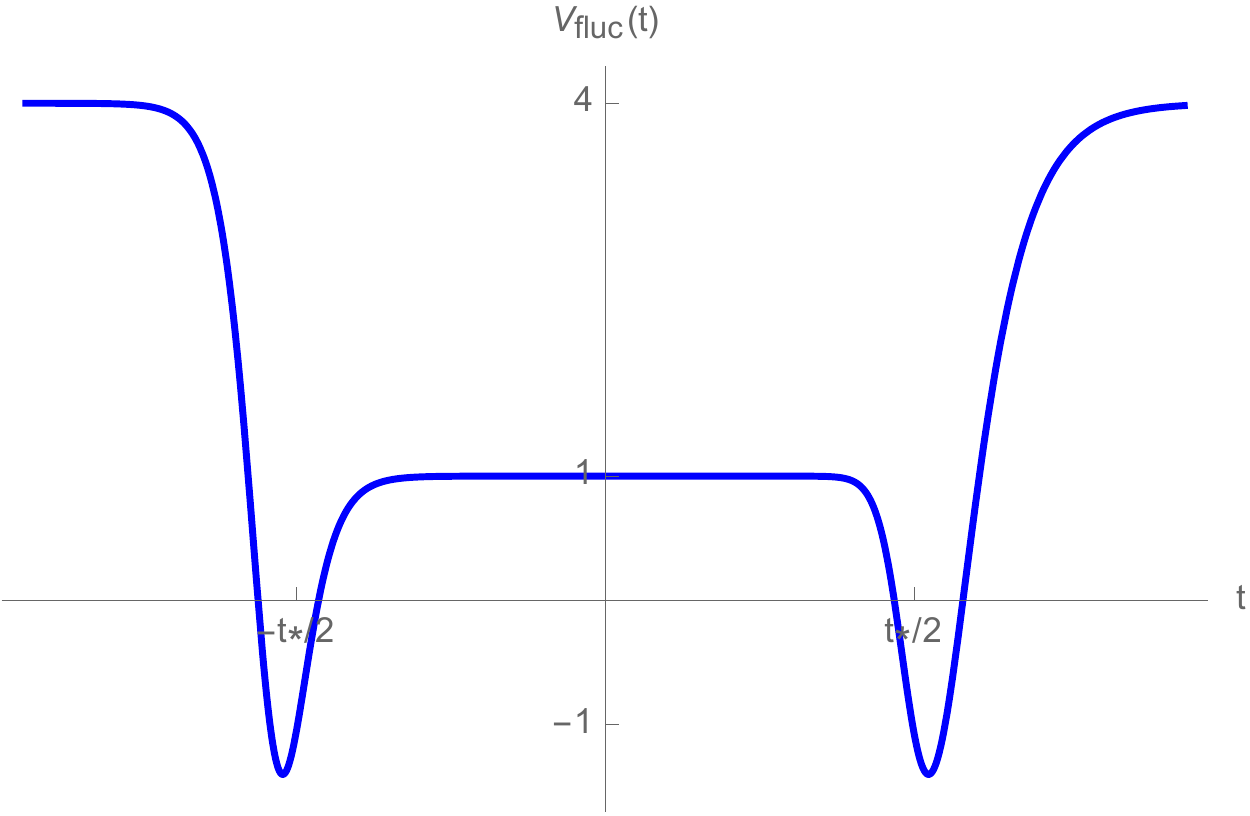}
\caption{Sketch of the structure of the fluctuation potential for the bion configurations.}
\label{fig:bion-fluc}
\end{figure}

\subsubsection{Physics of the Topological Bion}

The non-perturbative effect of the topological bion is analogous to that of a single instanton in the symmetric double-well potential in the sense that it is responsible for the non-perturbative splitting, $\Delta E\sim e^{-2S_I}$, of the energy levels for low-lying states localized in the outer wells. The prefactor is computed from the determinant factors as in expression (\ref{eq:inst-amp}), but now with action $2S_I$, and with the fluctuation potential of the form in Figure \ref{fig:bion-fluc}. Since the asymptotic values of the fluctuation potential coincide, the prefactor is finite and non-zero, unlike for the single instanton where it vanishes (\ref{eq:inst-det}):
\begin{eqnarray}
\text{topological bion amplitude:} \quad \left[\bar{I}_{\pm}\, I_{\mp}\right] \sim  e^{-2S_I}
\label{eq:real-bion-amp}
\end{eqnarray}

\subsubsection{Physics of the Neutral Bion}

The neutral bions play a very different role. Their non-perturbative character is  similar to that of a correlated instanton/anti-instanton molecule  in the symmetric double-well system \cite{Behtash:2015zha,Behtash:2015loa,Behtash:2018voa,Fujimori:2017oab}. The associated amplitude can be computed in an instanton gas picture as follows. Consider a widely separated (separation much greater than the instanton size scale) instanton/anti-instanton molecule, interacting via the effective   potential 
\begin{eqnarray}
V_{\rm eff} (\tau)=  - \omega_1 c_1^2\, e^{-\omega_1\, \tau }
  +\frac{\hbar}{2}(\omega_1-  \omega_0)\, \tau
  \label{eq:eff-pot}
\end{eqnarray}
where we write $\omega_0=\omega$ for the frequency of the inner well, and $\omega_1=2\omega$ for that of the outer wells. The first term in $V_{\rm eff} (\tau)$ is the classical interaction, arising from the overlap of the tails of two consecutive instantons, as is familiar from the usual symmetric double-well potential. \cite{Bogomolny:1980ur,ZinnJustin:1981dx,Behtash:2015zha,Behtash:2015loa,Behtash:2018voa,Fujimori:2017oab}
  The second term is the action accumulated during the intermediate Euclidean time regime spent in the other minimum. The difference in  energy between the true and false vacua is $\Delta E=  \frac{1}{2}(\omega_1-  \omega_0) $, and the time spent is $\tau$. Hence the action cost of this intermediate regime is $\Delta S= \Delta E \tau= \frac{\tau}{2}(\omega_1-  \omega_0)$. So for the symmetric triple-well potential in (\ref{eq:vdstw}), $\Delta S= \frac{\tau}{2}\omega$.

Unlike instantons, whose amplitude vanishes at the one-loop level (since the time spent in the false vacuum is infinite), the amplitude for a neutral bion is non-zero. 
To see this, consider a correlated  instanton/anti-instanton  pair, starting in the central well at the true vacuum with  frequency $\omega$, and taking a journey to the neighboring degenerate minimum with frequency $\omega_1 > \omega_0$, spending Euclidean  time $\tau$, (which is the separation quasi-zero mode that we will integrate over), and returning back to the original vacuum. 
To account for this correlated event, we evaluate the $\left[I_{+}\, \bar I_{+}\right]$ amplitude as follows
\begin{eqnarray}
\left[I_{+}\, \bar I_{+}\right] \sim e^{-2S_I}\int_{\Gamma_{\rm QZM}} d (\omega_1 \tau)  
 e^{ - \frac{1}{\hbar} \left(   - \omega_1 c_1^2\, e^{-\omega_1\tau }
  +\frac{\hbar}{2}(\omega_1-  \omega_0) \tau 
    \right) }
    \end{eqnarray} 
where the integral is to be taken along an appropriate  thimble \cite{Behtash:2018voa}, which, for $\beta\rightarrow\infty$, passes along the $\text{Im }\tau=\pi$ line.  This integral encodes the effect of the neutral bion fluctuation determinant. 

The critical point of the effective potential $V_{\rm eff} (\tau)$ determines the characteristic size of the neutral bion: 
\begin{eqnarray}
V_{\rm eff}^{'} (\tau)&=0 \Rightarrow 
\omega_1\tau^*  = \log  \left[ \frac{\omega_1 c_1^2}{\hbar {\cal D}} \right] \mp i \pi
\end{eqnarray}
We have defined  the ``deficit parameter'' 
\begin{eqnarray}
{\cal D}=  \frac{1}{2}\left( 1 -\frac{\omega_0}{\omega_1}\right)
\label{eq:deficit1}
\end{eqnarray}
The integration $\Gamma_{\rm QZM} $ over the quasi-zero-mode degree of freedom  yields 
\begin{eqnarray}
 \left[ I_{+}\, \bar I_{+}\right] _{\pm} \sim e^{-2S_I} e^{\pm  i \pi {\cal D} }  \left( \frac{\hbar}{\omega_1 c_1^2} \right)^{\cal D} \Gamma( {\cal D} ) 
\end{eqnarray}
This result identifies the hidden topological angle (HTA) \cite{Unsal:2012zj,Behtash:2015kna} as
\begin{eqnarray}
\theta_{{\rm HTA}}=  \frac{\pi}{2}\left(\frac{\omega_1 - \omega_0}{\omega_1}\right)  \equiv  \pi {\cal D}
\label{HTA-a}
\end{eqnarray}
In general, there would be a different HTA for a neutral bion connecting with another neighboring minimum with a different characteristic frequency. But in the symmetric triple-well potential (\ref{eq:vdstw}) the frequencies of the two outer wells are equal, so the amplitude for the different types of neutral bions are the same. Since $\omega_0=\omega$ and $\omega_1=2\omega$, we obtain a hidden topological angle given by:
\begin{eqnarray}
\theta_{{\rm HTA}}=  \frac{\pi}{4}
\label{HTA2}
\end{eqnarray}
Hence the quantum neutral bion amplitude is 
\begin{eqnarray}
 [ I_{+}\bar I_{+}] _{\pm} \sim e^{-2S_I} e^{\pm  i \frac{\pi}{4}  }  \left( \frac{\hbar}{\omega c_1^2} \right)^{1/4} \Gamma\left(\frac{1}{4}\right) 
\end{eqnarray}
 A similar discussion applies to all neutral bion amplitudes: $[ I_{\pm}\bar I_{\pm}]$ and $[ \bar I_{\pm} I_{\pm}]$.
Note that these amplitudes have an imaginary  two-fold ambiguity. The ambiguity cancels the ambiguity associated with the Borel resummation of perturbation theory \cite{Bogomolny:1980ur,ZinnJustin:1981dx,Behtash:2015zha,Behtash:2015loa,Behtash:2018voa,Fujimori:2017oab}, and the real part provides a non-perturbative shift to the ground state energy. 

In Section \ref{sec:zeta} we consider a deformation of the triple-well system in which the hidden topological angles differ by special fractionally-quantized amounts. This has the effect that perturbation theory is convergent for a finite set of states, but divergent for all other states.

\section{Supersymmetric Extension of the Triple Well System}
\label{sec:susy}

\subsection{SUSY Symmetric Triple Well}
\label{sec:susy-tw}

\begin{figure}[ht]
\centering
\includegraphics[width=0.5\textwidth]{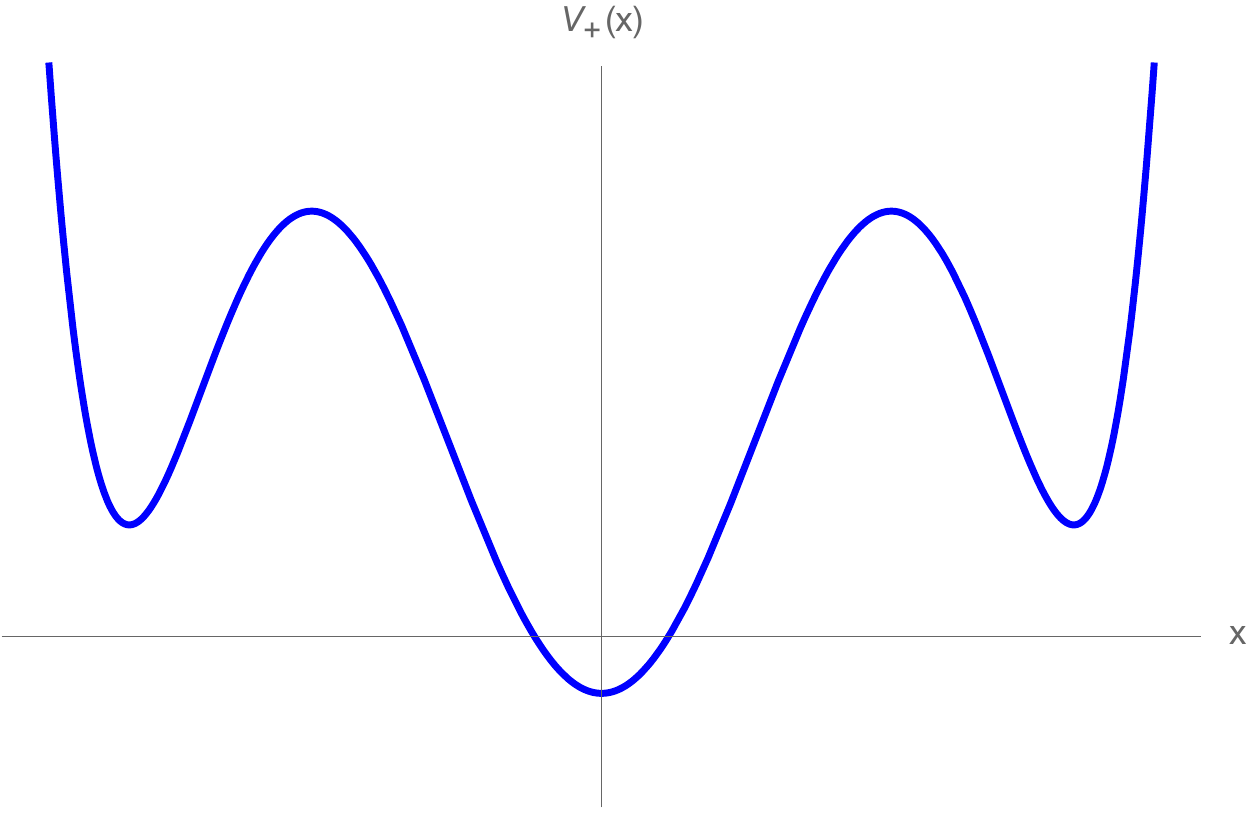}\label{fig:vpma}
\includegraphics[width=0.5\textwidth]{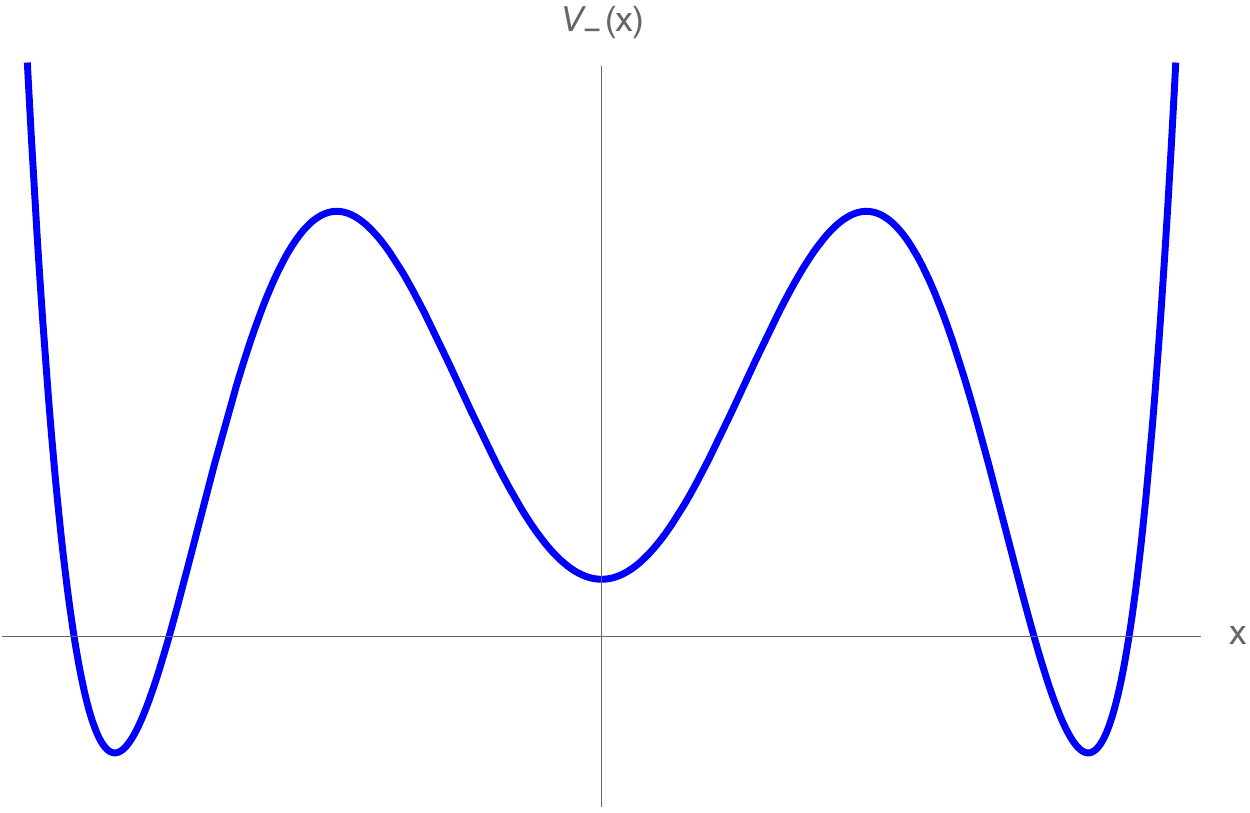}\label{fig:vpmb}
\caption{The form of the SUSY partner potentials $V_\pm(x)$  for the SUSY triple-well system.} \label{fig:vpm} 
\end{figure}

Integrating out the fermions in SUSY quantum mechanics produces a pair of bosonic partner Hamiltonians \cite{Witten:1981nf,Balitsky:1985in,Behtash:2015zha,Behtash:2015loa,Behtash:2018voa}:
\begin{eqnarray}
H^{\pm}=-\frac{\hbar^2}{2}\frac{d^2}{dx^2}+\frac{\omega^2}{2}x^2(x^2-1)^2\pm \frac{\hbar\,\omega}{2}(3\,x^2-1)
\label{eq:hpm}
\end{eqnarray}
The potential has a  classical term  and a quantum induced $O( \hbar)$ term  coming from integrating out the fermions. These partner Hamiltonians can be factored as
\begin{eqnarray}
H^\pm=\frac{1}{2} Q^\pm\, Q^\mp\qquad, \quad Q^\pm \equiv \pm \hbar\, \frac{d}{dx}+W^\prime
\label{eq:qq}
\end{eqnarray} 
where the superpotential term is $W^\prime(x)=\omega\, x(x^2-1)$. The associated partner  potentials, $V_\pm(x)=\frac{1}{2}(W^\prime)^2\pm \frac{\hbar}{2} W^{\prime\prime}$, are shown in Figure \ref{fig:vpm}. Note that it is important that the magnitude of the fermionic contribution, $\hbar\, W''(x)$ is parametrically small, proportional to $\hbar$, but still much greater than $e^{-2/\hbar}$, at which point the non-perturbative effects of the non-SUSY system dominate the physics.

As expected, the ground state of each potential $V_\pm$ has zero energy to all orders in perturbation theory. All other states have divergent perturbative expansions. (Nevertheless, there are still resurgent relations between perturbative and non-perturbative sectors \cite{Dunne:2016jsr}).
The perturbative structure of the SUSY partner hamiltonians $H^\pm$ can be studied with the BenderWu Mathematica package \cite{Sulejmanpasic:2016fwr}. This enables one to verify explicitly the generalization of the relation (\ref{eq:pt-relation}) between the perturbative expansions in the inner and outer wells extended to the SUSY potentials. We find that (recall $B\equiv \nu+\frac{1}{2}$, where $\nu$ is the perturbative level number for states localized in that well) the perturbative expansions for states localized in the inner and outer wells are related as:
\begin{eqnarray}
E_{\rm pert}^{\rm outer, \pm}(B, \hbar)&=&E_{\rm pert}^{\rm inner, \pm}\left(2B\pm\frac{3}{2}, \hbar\right)
\label{eq:susy-pt-relation}
\end{eqnarray}
This is easy to check using the BenderWu package.\footnote{To check this one can use the \texttt{BenderWuLevelPolynomial} function to compute the level-number dependence. See the documentation of the BenderWu package.\cite{Sulejmanpasic:2016fwr}}
\begin{figure}[htbp] 
 \vspace{-1cm}    \centering
   \includegraphics[width=\textwidth]{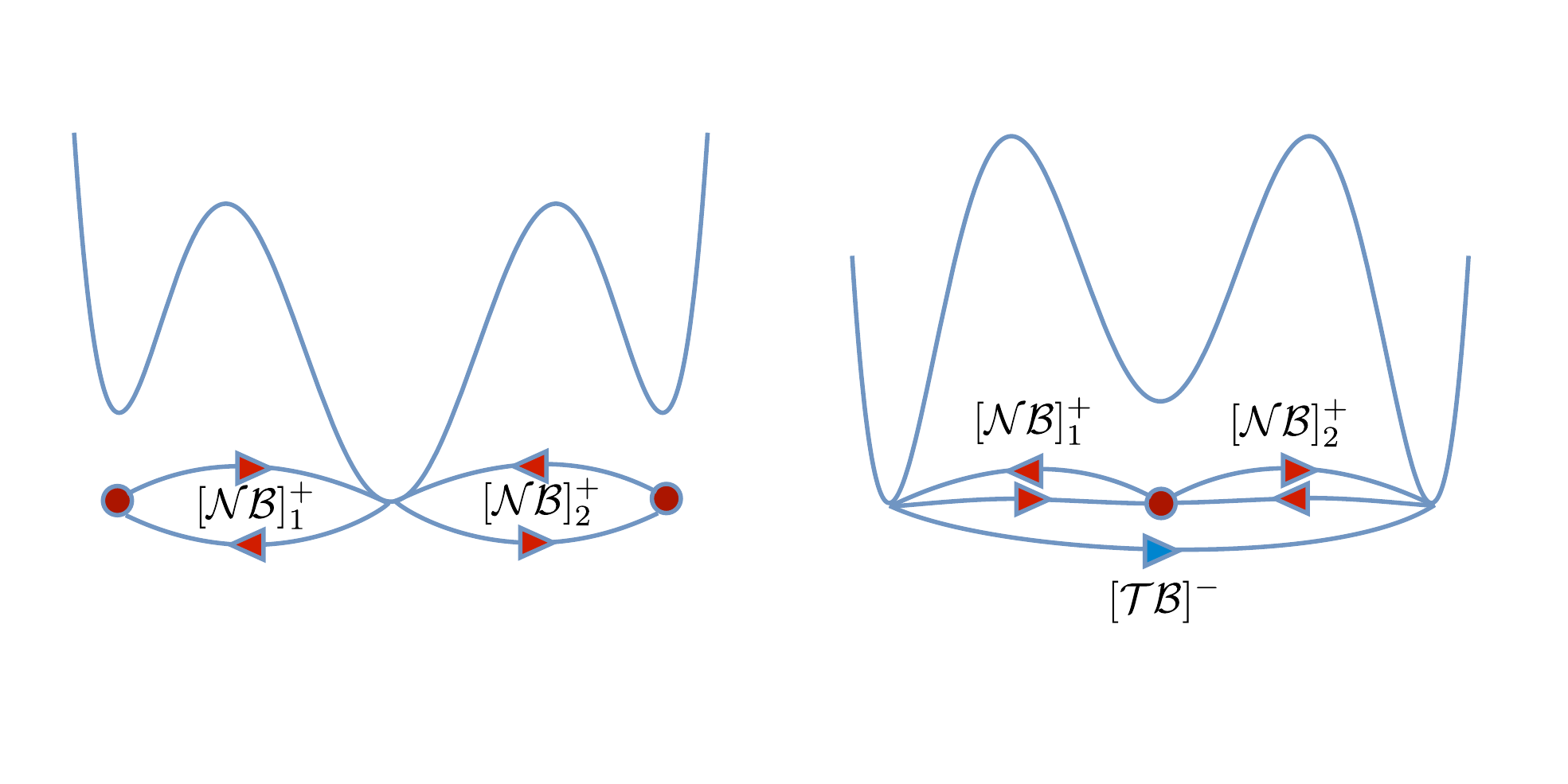}
   \vspace{-2cm} 
   \caption{Sketch of the neutral and topological bions in the $H^+$ (left) and $H^-$ (right) sectors.}
   \label{fig:susy-cancel}
\end{figure}

The semiclassical analysis of the SUSY triple-well system can be described as follows.
Recall that near the classical minima $x=0,\pm 1$, the bosonic potential behaves as
\begin{eqnarray}
V_{\rm bosonic}\equiv \frac{1}{2}W'(x)^2\approx\left\{\begin{array}{ll} 
\frac{4\omega^2}{2}x^2\quad, & \quad x\approx \pm 1\\
\frac{\omega^2}{2}x^2\quad, & \quad x\approx 0
 \end{array}  \right.
 \label{eq:vbos}
\end{eqnarray}
The classical harmonic oscillator frequencies are $\omega_0=\omega$, and $\omega_{1}=2 \omega$. Therefore a classical path fixed near $x=0,\pm 1$ will contribute
$\beta\omega_{0,1}/2$ to the action density, to one loop in perturbation theory in $\hbar$. However, such paths also get a contribution from the fermionic potential, $W''(x)$:
\begin{eqnarray}
\pm \frac{\hbar}{2}W''(x)\approx\left\{\begin{array}{ll} 
\pm \frac{2\hbar\omega}{2}\quad ,& \quad x\approx \pm 1\\
\mp\frac{\hbar\omega}{2}\quad ,& \quad x\approx 0
 \end{array} \right.
\end{eqnarray}
Thus, for $H^+$ the action contribution for paths near $x\approx 0$ cancel between the one loop bosonic fluctuation and the fermionic contribution, but they add near $x\approx \pm 1$. On the other hand, for $H^-$ the action contributions for paths near $x\approx \pm 1$ cancel, while they add near $x\approx 0$. In fact one can show that this cancellation is exact to any order, and so the energies in the corresponding local minima are all zero to any order in perturbation theory.

Thus, for the ground state a non-perturbative path must start and end at $x=0$ for the Hamiltonian $H^+$, but it must start and end  at $x=\pm 1$ for $H^-$.  See Figure \ref{fig:susy-cancel}. Therefore, $H^+$  allows only an instanton $I_{\pm}$ followed by an anti-instanton $\bar I_{\pm}$, i.e. a correlated $[I_{\pm}\bar I_{\pm}]$ configuration, interpolating from $x=0$ to $x=\pm 1$ and then back to $x=0$. We refer to these configurations as  \emph{neutral bions}, because their path is topologically trivial in the sense that they start and end at the same perturbative vacuum. The $H^-$ case also allows for a neutral bion: i.e. an instanton $I_{\pm}$ followed by an anti-instanton $\bar I_{\pm}$. However since $H^-$  has two degenerate minima to all orders in perturbation theory,
we also have a topologically nontrivial configuration: an anti-instanton of type $\bar I_{\pm}$, followed by an instanton of type $I_{\mp}$, thereby taking $x:\pm 1\rightarrow 0\rightarrow \mp 1$. We refer to such a  configuration as a \emph{topological bion}.

The correlated semiclassical configurations which contribute to the partition function are therefore:
\begin{eqnarray}
\text{neutral bion:}\quad &[\mathcal{NB}]^+\equiv [I_{\pm} \bar I_{\pm}]\quad &\text{for $H^+$}\\ 
\text{neutral bion:}\quad &[\mathcal{NB}]^-\equiv [I_{\pm} \bar I_{\pm}]\quad &\text{for $H^-$}
\\
\text{topological bion:}\quad &[\mathcal{TB}]^-\equiv[\bar I_{\pm} I_{\mp}]\quad &\text{for $H^-$}
\end{eqnarray}
These configurations are parametrized by the distance $\tau$ between the constituent instanton and anti-instanton. To compute their effect, one  integrates over the separation $\tau$: 
\begin{eqnarray}
&[I_{\pm} \bar I_{\pm}]=C_+e^{-2S_I}\int d\tau e^{-S_{nb}^+(\tau)- \omega_1 \tau}\quad, & \text{for $H^+$}\\ 
&[ I_{\pm}\bar I_{\pm}]=C_-e^{-2S_I}\int d\tau e^{-S_{nb}^-(\tau)- \omega_0 \tau}\quad, & \text{for $H^-$}\\
&[\bar I_{\pm} I_{\mp}]=C_-e^{-2S_I}\int d\tau e^{-S_{tb}^-(\tau)- \omega_0 \tau}\quad, & \text{for $H^-$}
\end{eqnarray}
The constants $C_{\pm}$ factorize to a pair of one-loop determinants around single (anti-)instantons, provided that the divergent part due to the frequency mismatch is subtracted, as it is explicitly taken into account above. The difference between $C_+$ and $C_-$ is determined entirely by the difference of the terms $\pm \int W''(x)$ in the action, evaluated in the background of the relevant configurations. Furthermore $S_{nb}(\tau)$ and $S_{tb}(\tau)$ are the ``classical'' (i.e. leading order in $\hbar$) interactions, and the subscript refers to the ``neutral bion'' and ``topological bion'', respectively. As usual, these interaction terms can be found from the asymptotic behavior of the bion solutions \cite{Behtash:2015zha,Behtash:2015loa,Behtash:2018voa}. A straightforward computation yields: 
\begin{eqnarray}
S_{nb}^+(\tau)&=&-2\omega_1 c_1^2 e^{-\omega_1\tau}\;,\\ 
S_{nb}^-(\tau)&=&-2\omega_0  c_0e^{-\omega_0 \tau}\;,\\
S_{tb}^-(\tau)&=& 2\omega_0  c_0e^{-\omega_0 \tau}= -S_{nb}^-(\tau)
\label{interactions}
\end{eqnarray}
where the coefficients, $c_0= 1/\sqrt 3 $ and $c_1= -3/2$, come from the details of the asymptotics of the instanton solution. 
Note that, as expected, for the neutral bion the classical interaction between constituents is attractive, while for the topological bion it is repulsive. Furthermore, also as expected, in the $H^-$ system the interaction between the topological bion constituents is minus that of the neutral bion constituents.

As explained in Sec. \ref{sec:dstw-bions}, these bions are the dominant configurations of a critical point at infinity.\cite{Behtash:2018voa} 
The saddle  point at infinity is a non-Gaussian critical point with vanishing contribution, but its thimble contributes non-trivially.  The evaluation over the thimble amounts to integrating over a contour  $\omega_1 \tau  \in {\mathbb R} + i \pi$,  $\omega_0 \tau  \in {\mathbb R} + i \pi$ in the first and  second cases (neutral bions), and $\omega_0 \tau  \in {\mathbb R}$ in the third case (topological bions).  
Performing the   $\tau$ integrals  {\it exactly} yields the amplitude and also the phase associated with each bion configuration:
\begin{eqnarray}
\text{$H^+$}\,:\quad [\mathcal{NB}]^+&\equiv& [I_{\pm} \bar I_{\pm}]=\frac{\hbar}{\omega_0\omega_1 c_0 c_1}e^{-2S_I + i \pi }\\ 
\text{$H^-$}\,:\quad [\mathcal{NB}]^-&\equiv& [ I_{\pm}\bar I_{\pm}]=\frac{\hbar}{\omega_0 \omega_1  c_0 c_1}e^{-2S_I + i \pi}\\
\text{$H^-$}\,:\quad  [\mathcal{TB}]^-&\equiv& [\bar I_{\pm} I_{\mp}]=\frac{\hbar}{\omega_0\omega_1  c_0 c_1}e^{-2S_I}
\end{eqnarray}
The phase is given by $\theta_{\rm HTA}= \pi$ both for the neutral bion $[\mathcal{NB}]^+$ in the $H^+$ sector, and for the neutral bion $[\mathcal{NB}]^-$ in the $H^-$ sector. In contrast, the phase associated with the topological bion is zero. The quantization of the HTA in units of $\pi$ for the neutral bion implies that there is no ambiguity in the neutral bion amplitude, and hence there should not be any ambiguity in perturbation theory for the ground state for $H^-$. Indeed, perturbation theory for the ground state is convergent in either sector (in fact, it vanishes to all perturbative orders due to supersymmetry).

\subsection{Non-perturbative Cancellations and the Witten index} 
\label{sec:witten}

We can now combine the contributions of the neutral and topological bions. Consider first the $H^+$ sector. In the leading order semi-classical approximation, we sum over the neutral bion configurations. There are two distinguishable configurations of equal magnitude which contribute to the partition function (see Fig. \ref{fig:susy-cancel}). The partition function can be viewed as a dilute gas of neutral bions of these two types. Summing over all such configurations, 
we find 
\begin{align}
Z^+ &\approx \sum_{n=0}^{\infty} \frac{1}{n!} (\beta[\mathcal{NB}])^n \sum_{m=0}^{\infty} \frac{1}{m!} (\beta[\mathcal{NB}])^m  \cr
&= e^{2\beta[\mathcal{NB}]} =  e^{-2\beta Ke^{-2S_I}}
\end{align}
with  $K$ a positive constant.
The non-perturbative ground state ($N=0$) energy in the $H^+$ sector is 
\begin{eqnarray}
E^{\rm np,+}(N=0) = - 2 \left[\mathcal{NB}\right]= -2 K e^{-2S_I + i \pi} >0
\label{eq:np+}
\end{eqnarray}
Here the label $N$ refers to all states in the spectrum, not just the perturbative levels in one of the potential wells.
Note that the positive semi-definiteness of the spectrum [guaranteed by the SUSY factorization (\ref{eq:qq})] arises semiclassically in (\ref{eq:np+}) as a result of the  hidden topological angle: $\theta_{HTA}=\pi$.  The HTA arises here because the integration cycle is in the complex domain, hence, the neutral bion configuration (or any other configuration on this thimble that contributes to semi-classics at this order) is manifestly complex. Since the contribution of real saddles or real configurations to path integrals are manifestly negative, the existence of   $\theta_{HTA}=\pi$ is strictly necessary in this case for compatibility with the supersymmetry algebra, which requires $E^{\rm np,+}(N=0) \geq 0$.

For the $H^-$ sector we  sum over both the neutral bions and also the topological bion. However the topological bion starts in one vacuum and ends in the other, so only even powers of these configurations can contribute to the thermal partition function [which demands periodicity in Euclidean time]. 
In addition there is a neutral bion, the sum over which is not constrained. Therefore, 
\begin{align}
Z^- &\approx 2 \sum_{n=0}^{\infty} \frac{1}{n!} (\beta[\mathcal{NB}])^n  
\sum_{m=0 }^{\infty} \frac{1}{(2m)!} (\beta[\mathcal{TB}])^{2m}     \cr
 &=2\cosh([\mathcal{TB}]\beta)e^{[\mathcal{NB}]\beta}\;.
\label{TPF}
\end{align}
The overall factor of two in front is due to the fact that there is a sum over the two perturbative supersymmetric vacua, i.e. the outer well harmonic vacua. However, notice that since $[\mathcal {NB}]^+=[\mathcal{NB}]^-=-[\mathcal{TB}]^-=-Ke^{-2S_I}$, where $K$ is a (positive) constant, we have that
\begin{eqnarray}
&Z^-\approx 1+e^{-2\beta K e^{-2S_I}}\;.
\end{eqnarray}
This has the consequence that the non-perturbative energies of the two lowest levels in the $H^-$ sector are
\begin{align}
&E^{\rm np, -}(N=0) = -(  [\mathcal{TB}]  +    [\mathcal{NB}])  =  0  \cr
&E^{\rm np, -}(N=1) =  -(  - [\mathcal{TB}]  +    [\mathcal{NB}]) =  2 Ke^{-2S_I}
\end{align}
Observe that $E^{\rm np, -}(N=1)$ is degenerate with $E^{\rm np,+}(N=0)$, as we know from supersymmetry.\cite{Witten:1981nf}
The supersymmetric Witten index is therefore given semiclassically by
\begin{eqnarray}
I_W=Z^--Z^+=1
\end{eqnarray}
which agrees with the well known result from SUSY.\cite{Witten:1981nf}
Semiclassically, the cancellation of the ground state energy is due to the opposite sign contributions of $ [\mathcal{TB}]$ and   $[\mathcal{NB}]$, which arises from the hidden topological angle $\theta_{\rm HTA}=\pi$.\cite{Behtash:2015zha} 

\section{Zeta-deformed Theories and Quasi-Exactly Soluble Models}
\label{sec:zeta}

The fermionic contribution to the partner Hamiltonians (\ref{eq:hpm}) can be extended to multiple flavors of fermions, where we will also take the ''fermion number'' parameter $\zeta >0$ to be non-integer, constructing $H^{\pm, \zeta}$ pairs of partner Hamiltonians.\cite{Balitsky:1985in,Verbaarschot:1990ga,Verbaarschot:1990fa,Behtash:2015loa}   In this section, we examine novel properties of these paired Hamiltonians:
\begin{eqnarray}\label{eq:zeta_H}
H^{\pm, \zeta}=-\frac{\hbar^2}{2}\frac{d^2}{dx^2}+\frac{\omega^2}{2}x^2(x^2-1)^2\pm \zeta \frac{\omega}{2}\hbar (3x^2-1)
\label{eq:hzeta}
\end{eqnarray}
Notice that the form mimics the supersymmetric form of the potential 
\be
V^{\pm, \zeta}(x)=\frac{1}{2}W'(x)^2\pm \frac{\zeta}{2} \,\hbar\, W''(x)
\ee
with $W'(x)$ being analogous to the ``superpotential''
\be
W'(x)=\omega\, x(x^2-1)
\ee

\subsection{Perturbative Expansions for the Zeta-deformed Theories}
\label{sec:pert-zeta}

We first observe that the perturbative expansion around the inner and outer wells are once again related. We employ the BenderWu mathematica package to facilitate these expansions, and discover that the energy expansion coefficients around the inner and outer wells, as a function of the harmonic oscillator level number, given in terms of $B=\nu+\frac{1}{2}$ for the respective well, are given by
\begin{eqnarray}
E^{{\rm inner},\pm,\zeta}&:& \left\{B\mp\frac{\zeta}{2},-\frac{3 B^2}{2}\pm\frac{3 B \zeta}{2}-\frac{3}{8},\right.\nonumber\\ 
&&\left.-3 B^3\pm\frac{9 B^2\zeta}{2}-\frac{9 B \zeta^2}{8}-\frac{21 B}{8}\pm\frac{9 \zeta}{8}, \dots\right\}\\
E^{{\rm outer},\pm,\zeta}&:& \left\{2 B\pm \zeta,-6 B^2\mp 6 B \zeta-\frac{9 \zeta^2}{8}-\frac{3}{8},\right. \nonumber\\ 
&&\left.-24 B^3\mp 36 B^2 \zeta-\frac{63 B
   \zeta^2}{4}-\frac{21 B}{4}\mp\frac{27 \zeta^3}{16}\mp\frac{45 \zeta}{16},\dots\right\}
   \end{eqnarray}
These expansion coefficients imply that the perturbative energies in the inner and outer wells are related as follows: 
\be\label{eq:zeta_in_out}
E^{{\rm outer},\pm,\zeta}_{{\rm pert}}(B,\hbar)=E^{\rm{inner,}\pm,\zeta}_{\rm pert}\left(2B\pm \frac{3\zeta}{2},\hbar\right)\;.
\ee
These relations  generalize the inner-outer perturbative relations for the non-SUSY case in (\ref{eq:pt-relation}), and for the SUSY case in (\ref{eq:susy-pt-relation}).

There is further interesting structure in the perturbative expansions for the partner hamiltonians $H^{\pm, \zeta}$, which we now use to give a semiclassical understanding of the rich algebraic structure of quasi-exactly-solvable (QES) hamiltonians.\cite{Turbiner:1987nw,Turbiner:1987kt,Turbiner:1994gi,itep,Kozcaz:2016wvy}. After suitable rescaling, the sextic potential analyzed  in Ref \cite{Turbiner:1994gi} matches the form in (\ref{eq:hzeta}). In our notation, the QES systems arise for  special rational values of $\zeta$:
\begin{eqnarray}
\zeta_{\rm QES} =\frac{2m+1}{3}\qquad, \quad m=1, 2, 3, \dots
\label{eq:qes-zeta}
\end{eqnarray}
Using the BenderWu package, we have studied the perturbative expansions for the low-lying states of the $H^{\pm, \zeta}$ partner hamiltonians. Note that for $H^{+, \zeta}$ the middle well is lowered, while for $H^{-, \zeta}$ the outer wells are symmetrically lowered, analogous to Figure \ref{fig:vpm} for the SUSY models.

\subsubsection{Perturbative Structure for $H^{+, {\zeta_{\rm QES}}}$}
\label{sec:qes+}

For $H^{+, {\zeta_{\rm QES}}}$ the middle well is lowered, so the low-lying states are localized in the middle well for $0\leq \nu_{\rm inner}\leq m-1$.
\begin{itemize}
\item
\underline{m=1}: The lowest QES case has $m=1$, which means $\zeta=1$, which is the SUSY case. The ground state has $E_{\rm pert}(\nu_{\rm inner}=0, \hbar)=0$ to all orders of perturbation theory. But it receives a non-perturbative shift, which is positive, as discussed in Section \ref{sec:witten}. For $\nu_{\rm inner}\geq 1$, the perturbative expansions are all divergent asymptotic expansions, with non-alternating expansion coefficients that grow factorially in magnitude.
\begin{subequations}
\begin{eqnarray}
\nu_{\rm inner}=0 &:& \text{convergent expansion}\quad E_{\rm pert}=0 \\
\nu_{\rm inner}\geq 1 &:& \text{divergent non-alternating expansion} 
\label{eq:m1+}
\end{eqnarray} 
\end{subequations}

\item
\underline{m=2}: In this case the deformation parameter has a non-integer rational value, $\zeta=\frac{5}{3}$, and we find that the ground state, $(\nu_{\rm inner}=0)$, has a divergent and non-alternating perturbative expansion, while the series for the first excited state truncates $(\nu_{\rm inner}=1)$, and is thus convergent:
\begin{subequations}
\begin{eqnarray}
\nu_{\rm inner}=0 &:& \text{divergent non-alternating expansion} \\
\nu_{\rm inner}=1 &:& \text{convergent expansion}\quad E_{\rm pert}=\frac{2}{3} \\
\nu_{\rm inner}\geq 2 &:& \text{divergent non-alternating expansion}
\label{eq:m2+}
\end{eqnarray} 
\end{subequations}

\item
\underline{m=3}:  In this case the deformation parameter has a non-integer rational value, $\zeta=\frac{7}{3}$, and we find the following pattern:
\begin{subequations}
\begin{eqnarray}
\nu_{\rm inner}=0 &:& \text{convergent expansion}\quad E_{\rm pert}=\frac{1}{3}-\sqrt{1-2\hbar} \\
\nu_{\rm inner}=1 &:& \text{divergent non-alternating expansion} \\
\nu_{\rm inner}=2 &:& \text{convergent expansion}\quad E_{\rm pert}=\frac{1}{3}+\sqrt{1-2\hbar} \\
\nu_{\rm inner}\geq 3 &:& \text{divergent non-alternating expansion}
\label{eq:m3+}
\end{eqnarray}
\end{subequations}

\item
\underline{m=4}:  In this case the deformation parameter has an integer  value, $\zeta=3$, and we find the following pattern:
\begin{subequations}
\begin{eqnarray}
\nu_{\rm inner}=0 &:&  \text{divergent non-alternating expansion} \\ 
\nu_{\rm inner}=1 &:& \text{convergent expansion}\quad E_{\rm pert}=1-\sqrt{1-6\hbar} \\
\nu_{\rm inner}=2 &:&  \text{divergent non-alternating expansion} \\
\nu_{\rm inner}= 3 &:& \text{convergent expansion}\quad E_{\rm pert}=1+\sqrt{1-6\hbar} \\
\nu_{\rm inner}\geq 4 &:& \text{divergent non-alternating expansion}
\label{eq:m4+}
\end{eqnarray}
\end{subequations}

\end{itemize}

This pattern continues for higher $m$: the states with $0\leq \nu_{\rm inner}\leq m-1$ alternate between convergent and divergent perturbative expansions, and all states receive non-perturbative  corrections of the form: $\Delta E^{\rm NP}\sim e^{-2S_I}$.

\subsubsection{Perturbative Structure for $H^{-, {\zeta_{\rm QES}}}$}
\label{sec:qes-}

For $H^{-, {\zeta_{\rm QES}}}$ the outer wells are lowered symmetrically, so the low-lying states are localized in the outer wells for $0\leq \nu_{\rm outer}\leq m-1$. These states have an additional parity structure because of the parity symmetry between the two outer wells.

\begin{itemize}
\item
\underline{m=1}: The lowest QES case has $m=1$, which means $\zeta=1$, which is the SUSY case. 
The lowest perturbative state in each outer well has $E_{\rm pert}(\nu_{\rm outer}=0, \hbar)=0$ to all orders of perturbation theory. This perturbative level is split by non-perturbative effects. The parity-symmetric ground state remains zero, while the parity-antisymmetric first-excited state receives a positive non-perturbative shift, $\Delta E^{\rm NP}\sim e^{-2S_I}$, as discussed in Section \ref{sec:witten}.
 For $\nu_{\rm outer}\geq 1$, the perturbative expansions are all divergent asymptotic expansions, with non-alternating expansion coefficients that grow factorially in magnitude. 
\begin{subequations}
 \begin{eqnarray}
\nu_{\rm outer}=0 &:& \text{convergent expansion}\quad E_{\rm pert}=0 \\
\nu_{\rm outer}\geq 1 &:& \text{divergent non-alternating expansion} 
\label{eq:m1-}
\end{eqnarray} 
\end{subequations}

\item
\underline{m=2}: In this case the deformation parameter has a non-integer rational value, $\zeta=\frac{5}{3}$, and we find that the lowest perturbative level, $(\nu_{\rm outer}=0)$, has a convergent (indeed, truncating) expansion, 
  \begin{eqnarray}
 E_{\rm pert}^{-, \zeta=\frac{5}{3}}(\nu_{\rm outer}=0, \hbar)= -\frac{2}{3}  
 \label{eq:53_N0}
 \end{eqnarray}
This lowest perturbative level is split into a doublet by non-perturbative effects, with the lower (ground) doublet state receiving a negative shift, $\Delta E^{\rm NP}\sim - e^{-2S_I}$, while the higher (first excited) doublet state receives no non-perturbative shift.
The general perturbative pattern for $m=2$ is:
\begin{subequations}
\begin{eqnarray}
\nu_{\rm outer}=0 &:&  \text{convergent perturbative expansion} \\
\nu_{\rm outer}\geq 1 &:&  \text{divergent non-alternating expansion} 
\label{eq:m2-}
\end{eqnarray} 
\end{subequations}
\item
\underline{m=3}:  In this case the deformation parameter has a non-integer rational value, $\zeta=\frac{7}{3}$, and we find that the first two perturbative levels for states localized in the outer wells have non-truncating but {\it convergent} (indeed, exactly summable) expansions:
\begin{subequations}
  \begin{eqnarray}
 E_{\rm pert}^{-, \zeta=\frac{7}{3}}(\nu_{\rm outer}=0, \hbar)&=&-\frac{1}{3} - \sqrt{1+2\hbar} 
  \label{eq:73_N01a}\\
  E_{\rm pert}^{-, \zeta=\frac{7}{3}}(\nu_{\rm outer}=1, \hbar)&=&-\frac{1}{3}+ \sqrt{1+2\hbar} 
 \label{eq:73_N01b}\\
   E_{\rm pert}^{-, \zeta=\frac{7}{3}}(\nu_{\rm outer}\geq 2, \hbar)&:& \text{divergent non-alternating expansion} 
 \end{eqnarray}
 \end{subequations}
These should be contrasted with the $\nu_{\rm inner}=0, 2$ perturbative energies for the $H^{+, \zeta}$ sector with $\zeta=\frac{7}{3}$ in the previous subsection. The parity-symmetric forms of the $\nu_{\rm outer}=0, 1$ perturbative states in (\ref{eq:73_N01a}-\ref{eq:73_N01b}) are exactly solvable, and receive no non-perturbative corrections. On the other hand, the parity-antisymmetric forms of the $\nu_{\rm outer}=0, 1$ states in (\ref{eq:73_N01a}-\ref{eq:73_N01b}) receive non-perturbative corrections with a positive shift: $\Delta E^{\rm NP}\sim e^{-2S_I}>0$. For $\nu_{\rm outer}\geq 2$, the perturbative energy expansions are divergent with non-alternating coefficients growing factorially fast in magnitude.
\item
\underline{m=4}:  In this case the deformation parameter has an integer value, $\zeta=3$, and we find that the first two perturbative levels for states localized in the outer wells have non-truncating but {\it convergent} (indeed, exactly summable) expansions:
\begin{subequations}
  \begin{eqnarray}
 E_{\rm pert}^{-, \zeta=3}(\nu_{\rm outer}=0, \hbar)&=&-1- \sqrt{1+6\hbar} 
  \label{eq:3_N01a}\\
  E_{\rm pert}^{-, \zeta=3}(\nu_{\rm outer}=1, \hbar)&=&-1+ \sqrt{1+6\hbar} 
 \label{eq:3_N01b}\\
   E_{\rm pert}^{-, \zeta=3}(\nu_{\rm outer}\geq 2, \hbar)&:& \text{divergent non-alternating expansion} 
 \end{eqnarray}
 \end{subequations}
These should be contrasted with the $\nu_{\rm inner}=1, 3$ perturbative energies for the $H^{+, \zeta}$ sector with $\zeta=3$ in the previous subsection. The parity-symmetric forms of the $\nu_{\rm outer}=0, 1$ perturbative states in (\ref{eq:3_N01a}-\ref{eq:3_N01b}) are exactly solvable, and receive no non-perturbative corrections. On the other hand, the parity-antisymmetric forms of the $\nu_{\rm outer}=0, 1$ states in (\ref{eq:3_N01a}-\ref{eq:3_N01b}) receive non-perturbative corrections with a positive shift: $\Delta E^{\rm NP}\sim e^{-2S_I}>0$. For $\nu_{\rm outer}\geq 2$, the perturbative energy expansions are divergent with non-alternating coefficients growing factorially fast in magnitude.

\end{itemize}

\subsection{Semiclassical Bion Analysis of QES Spectra: the Hidden Topological Angle and the Discrete $\theta$ Angle}
\label{sec:qes-bion}

In this Section we present a bion explanation of the intricate spectral patterns found for the $H^{\pm, \zeta}$ QES systems in the previous Section.

\subsubsection{$H^{+, \zeta}$: Physics of the Hidden Topological Angle}
\label{sec:hta}

 The partner  Hamiltonian
$H^{+, \zeta}$ is the $\zeta$-deformation of the partner Hamiltonian in SUSY QM which does not have a normalizable zero energy state, with potential  shown in the left hand side of 
Fig.~\ref{fig:vpm}. 
For $H^{+, \zeta}$, the inner-well is lowered and the outer wells are lifted. 
The harmonic  states in the inner  well have energies  $\omega_0 \left(\frac{1}{2}  - \frac{\zeta}{2} + \nu_{\rm inner} \right)$, where  $\nu_{\rm inner}=0, 1, 2, \ldots$, and the lowest harmonic  state  in each of the outer wells has energy $\omega_1 \left(\frac{1}{2}  + \frac{\zeta}{2} \right)$.  Recall that for our sextic potential, $\omega_0=\omega$, and $\omega_1=2\omega$.
We wish to explain semiclassically the alternating pattern structure of the low lying states 
in the inner well found in Section \ref{sec:qes+}, for which for half of the states perturbation theory is convergent, while for the other half it is an asymptotic  divergent series, with non-alternating coefficients.  This arises due to an interesting structure of the HTA.  However, for the convergent states, the result will not converge to the physical non-perturbative answer, as there are only neutral bion contributions, and no topological bions to cancel them. Hence the non-perturbative effects are present, unlike for the symmetric states in the $H^{-, \zeta}$ system which we discuss below in the next subsection.

The semiclassical amplitude for the neutral bion configuration can be computed via a generalization of the method in Section \ref{sec:susy-tw}, extended to include the dependence on the perturbative level number $\nu$. This analysis builds on earlier work\cite{Verbaarschot:1990fa,ZinnJustin:2004ib,Behtash:2015loa}, and full details of the modern bion approach will be presented elsewhere. \cite{toappear} 
The  main result is that the thimble integration leads to the following expression for the neutral bion amplitude:
\begin{subequations}
\begin{eqnarray}
 [ I_{+}\bar I_{+}] _{\pm} \sim e^{-2S_I} e^{\pm  i \pi {\cal D}^{+, \zeta} }  \left( \frac{\hbar}{\omega_1 c_1^2} \right)^{{\cal D}^{+, \zeta} } \Gamma( {\cal D}^{+, \zeta} ) 
 \label{c-b-plus}
\end{eqnarray}
\end{subequations}
Here the  ``deficit angle'' ${\cal D}^{+, \zeta}$ is (compare with Eqs. (\ref{eq:deficit1}) and (\ref{eq:deficit2})):
\begin{eqnarray}
{\cal D}^{+, \zeta}=\frac{1}{2}  \left[    \left( 1+ \zeta\right)  - \frac{\omega_0}{\omega_1}  \left(1 - \zeta  +  2\,\nu_{\rm inner} \right)\right]
 \label{eq:deficit3}
\end{eqnarray}
The neutral bion amplitude has an imaginary ambiguous part
\begin{eqnarray}
{\rm Im}( [ I_{+}\bar I_{+}] _{\pm})  &\sim&  \pm i e^{-2S_I} \frac{\pi}{\Gamma(1-  {\cal D}^{+, \zeta} ) } 
  \left( \frac{\hbar}{\omega_0 c_0^2} \right)^{{\cal D}^{+, \zeta}}   
 \label{c-b-4}
\end{eqnarray}
For the symmetric triple-well, for which  $\omega_1= 2 \omega_0=2\omega$, 
we deduce the hidden topological angle:
\begin{align}
\theta^{+, \zeta}_{{\rm HTA}}=   \pi {\cal D}^{+, \zeta} =   \pi \left(  \frac{1}{4}  + \frac{3\zeta}{4}  - \frac{ \nu_{\rm inner}}{2}  \right)  = 
\frac{\pi}{2}\left( m+1-\nu_{\rm inner} \right)
\label{eq:HTA-zeta-3}
\end{align}
Thus, for the $H^{+, \zeta}$ sector, the HTA  is quantized in units of $\pi/2$, rather than in units of $\pi$, as arises for the $H^{-, \zeta}$ sector: compare with Eq. (\ref{eq:HTA-zeta-1}) for $\theta^{-, \zeta}_{{\rm HTA}}$.

The result (\ref{eq:HTA-zeta-3}) for $\theta^{+, \zeta}_{{\rm HTA}}$ has several important implications for the structure of 
the low-lying levels for which $0 \leq \nu_{\rm inner}  \leq m-1$. There are two distinct cases, depending on whether $m+1-\nu_{\rm inner}$ is even or odd.  

\begin{itemize}
\item
\underline{\bf  $m$ odd}: The   imaginary ambiguous part vanishes for $\nu_{\rm inner}  \in {\cal S}_1 = \{0, 2, \ldots, m-1\}$,  and it does not vanish for   $\nu_{\rm inner}   \in {\cal S}_2 = \{1,3, \ldots, m-2 \}$. 
 This comes about  because $\theta^{+, \zeta}_{{\rm HTA}  }$ is quantized in units of 
 $\pi$ for  ${\cal S}_1$, but $\theta^{+, \zeta}_{{\rm HTA}  }$ is quantized in odd multiples of $\pi/2$ for  ${\cal S}_2$. Therefore, perturbation theory must be convergent for ${\cal S}_1$,   and it must be divergent for ${\cal S}_2$. Furthermore, for the states  in ${\cal S}_1$, 
 since $\Delta E^{\rm NP} = - [\cal NB] $, the non-perturbative energy shift  is positive if the HTA is an odd integer multiple of $\pi$, and is negative if the HTA  is an even integer multiple of $\pi$.

\item
\underline{\bf  $m$ even}: The   imaginary ambiguous part vanishes for $\nu_{\rm inner}  \in {\cal S}_1 = \{ 1,3,  \ldots, m-1\}$,  and it does not vanish for   $\nu_{\rm inner}   \in {\cal S}_2 = \{0, 2,4, \ldots, m-2 \}$. 
 This comes about  because $\theta^{+, \zeta}_{{\rm HTA}  }$ is quantized in units of 
 $\pi$ for  ${\cal S}_1$, but $\theta^{+, \zeta}_{{\rm HTA}  }$ is quantized in odd multiples of $\pi/2$ for  ${\cal S}_2$. Therefore, perturbation theory must be convergent for ${\cal S}_1$,   and it must be divergent for ${\cal S}_2$. Note that this has the interesting implication that for the ground state, perturbation theory is asymptotic, while for the first excited state it is convergent.

 Furthermore, for the states  in ${\cal S}_1$, 
 \begin{align}
 \Delta E^{\rm NP} = - [{\cal NB}] =  -   e^{  i  \pi \left(  \frac{m+1}{2} -   \frac{ \nu_{\rm inner}}{2} \right) }
 2K e^{-2S_I}  
\end{align} 
Therefore, the non-perturbative energy shift  is positive if the HTA is an odd integer multiple of $\pi$, and it  is negative if the HTA is an even integer multiple of $\pi$. For level number $\nu_{\rm inner} \geq m$,  all the states have divergent asymptotic expansion.   

 \end{itemize}
 
\noindent These predictions of the neutral bion analysis explain the patterns found in Section \ref{sec:qes+} using the  Bender-Wu Mathematica package.\cite{Sulejmanpasic:2016fwr}

\subsubsection{$H^{-, \zeta}$:  Physics of the Hidden Topological Angle and the Discrete $\theta$ Angle}
\label{sec:hta-theta}

When $\zeta$ is close to unity, the deformation \eqref{eq:zeta_H} serves as a soft SUSY breaking deformation which reveals how the hidden resurgent structure of SUSY quantum mechanics is present, disappearing at the non-generic SUSY value of $\zeta=1$  \cite{Dunne:2016jsr}. Consider the role of the neutral bions in this $\zeta$-deformed theory.   The neutral bion configuration starts at an outer well, interpolates to the inner well, and then interpolates back again. At the harmonic level, 
the lowest state localized in the inner well has energy $\omega_0 \left(\frac{1}{2}  + \frac{\zeta}{2}\right)$, and the ground state and 
low lying states localized in the outer wells have energies $\omega_1 \left(\frac{1}{2}  - \frac{\zeta}{2} + \nu_{\rm outer} \right)$,  where 
$\nu_{\rm outer}=0, 1, 2, \ldots$ is the harmonic energy level label. For a certain number of low lying states localized in the outer wells, we will show that the perturbation theory is convergent. For half of these states, perturbation theory converges to the exact physical result, while for the other half there is a non-perturbative bion contribution on top of the convergent perturbative sum, generalizing the result of the supersymmetric model. 

As in the $H^{+, \zeta}$ sector discussed in the previous section, the semiclassical amplitude for the neutral bion configuration can be computed via a generalization\cite{toappear} of the method in Section \ref{sec:susy-tw}, extended to include the dependence on the perturbative level number $\nu$. The  main result is that the thimble integration leads to the following expression for the neutral bion amplitude:
\begin{eqnarray}
 [ I_{+}\bar I_{+}] _{\pm} \sim e^{-2S_I} e^{\pm  i \pi {\cal D}^{-, \zeta} }  \left( \frac{\hbar}{\omega_0 c_0^2} \right)^{{\cal D}^{-, \zeta} } \Gamma( {\cal D}^{-, \zeta} ) 
 \label{c-b}
\end{eqnarray}
Here ${\mathcal D^{-, \zeta}}$ is the zeta-deformed ``deficit angle'' (compare with Eq. (\ref{eq:deficit1}) and (\ref{eq:deficit3})):
\begin{eqnarray}
{\mathcal D^{-, \zeta}}\equiv \frac{1}{2}  \left[  1- \frac{\omega_1}{\omega_0}(1+ 2 \nu_{\rm outer}) + \zeta\left(1+ \frac{\omega_1}{\omega_0}\right)
 \right] 
 \label{eq:deficit2}
 \end{eqnarray}
The imaginary part of the neutral bion amplitude can be written as 
\begin{eqnarray}
{\rm Im}( [ I_{+}\bar I_{+}] _{\pm})  &\sim&  \pm i e^{-2S_I} \frac{\pi}{\Gamma(1-  {\cal D}^{-, \zeta} ) } 
  \left( \frac{\hbar}{\omega_0 c_0^2} \right)^{{\cal D}^{-, \zeta} }   
 \label{c-b-2}
\end{eqnarray}
Thus, for the symmetric triple-well, for which  $\omega_1= 2 \omega_0=2\omega$, 
we deduce the hidden topological angle associated with the neutral bion configuration:
\begin{align}
\theta^{-, \zeta}_{{\rm HTA}  }=   \pi {\cal D}^{-, \zeta} =   \pi \left(  -\frac{1}{2}  +   \frac{3\zeta}{2}  - 2 \nu_{\rm outer}  \right)  =   \pi \left(  m  - 2 \nu_{\rm outer}  \right) 
\label{eq:HTA-zeta-1}
\end{align}
Note that $\theta^{-, \zeta}_{{\rm HTA}}$ is quantized in integer units of $\pi$, rather than in integer units of $\frac{\pi}{2}$, as for $\theta^{+, \zeta}_{{\rm HTA}}$ in (\ref{eq:HTA-zeta-3}).
Furthermore, 
the phase $\theta^{-, \zeta}_{{\rm HTA}}$ does not have  a level number dependence, since the angles are identified by $2\pi$ shifts. This is in contrast to the  $H^{+, \zeta}$ sector, where there is  a non-trivial level number dependence  in (\ref{eq:HTA-zeta-3}), leading to an alternating convergent/divergent perturbative pattern for the low-lying states of the  $H^{+, \zeta}$ sector. 
For the states for which $ 0 \leq  \nu_{\rm outer} \leq  \left\lfloor  \frac{3 (\zeta- 1)}{4} \right\rfloor =  \left\lfloor  \frac{m-1}{2} \right\rfloor  $, the amplitude  of the 
topological bions and neutral bions are related by 
\begin{align}
   \left[\mathcal{NB}\right]= e^{ i \pi m} \left[\mathcal{TB}\right]
   \end{align}
for the QES $\zeta$ values in (\ref{eq:qes-zeta}).  
    
Each value of $\nu_{\rm outer}$ gives rise to two eigenstates of the  $H^{-, \zeta}$ hamiltonian, one symmetric combination ($N= 2\nu_{\rm outer}$)  and one anti-symmetric ($N=2\nu_{\rm outer}+1$),  where $N$ is the fundamental quantum 
number associated with $H^{-, \zeta}$.  Here the level label $N$ refers not to a perturbative level in a given well, but to the states of the whole potential. The states for which there is no non-perturbative contribution, i.e, the non-perturbative contribution cancels, is determined by an  interference pattern sourced by the
hidden topological angle $\theta_{\rm HTA}$ and by a discrete theta angle   $\theta_{\rm disc.}$.

The appearance of the discrete theta angle can be seen as follows. 
The triple-well potential has a parity symmetry, $P$.   Therefore, one can consider 
 two types of partition functions: ${\rm tr} \left(e^{- \beta H^{-, \zeta}}\right)$ and ${\rm tr}\left(P\, e^{- \beta H^{-, \zeta}}\right)$ (see also \cite{ZinnJustin:2004ib}). One can now gauge parity. By this one usually means the following  
 \begin{align}
 {\rm tr}  \left[\left({ \frac{1+  P}{2}}\right) e^{- \beta H^{-, \zeta}}\right] = \sum_{ N \in {\cal H}_{\rm P \,even }}   e^{-\beta E_{N} } 
 \end{align}
where  in the state sum  we sum over only parity even states, {i.e. gauging the parity is equivalent to this projection. However when gauging one has a choice to project to a state sum over only parity odd states, and we can identify this with turning on the discrete theta angle}\footnote{{Here the terminology \cite{Aharony:2013hda} is analogous to the one in periodic potentials. The discrete translation symmetry of a periodic potential can be gauged by projecting to a particular charge $\theta$ which is angle valued for a $\mathbb Z$ symmetry. This is the usual $\theta$ angle, or Bloch angle, for a particle on a circle.}}:
 \begin{eqnarray}  
 \theta_{\rm disc.}=\pi
 \label{eq:theta-disc}
 \end{eqnarray}  
 Then
 \begin{align}
 {\rm tr}  \left[\left( \frac{1-P}{2}\right) e^{- \beta H^{-, \zeta}}\right] = \sum_{ N \in {\cal H}_{\rm P \,odd }}   e^{-\beta E_{N} } 
\end{align}
On the other hand, 
\begin{eqnarray}
Z^-={\rm tr}\left( e^{- \beta H^{-, \zeta}} \right) & \approx&       e^{- \beta E_{0} } 2\cosh([\mathcal{TB}]\beta)\, e^{[\mathcal{NB}]\beta} \nonumber\\
&=&
  e^{- \beta E_{0}}   \left(  e^{  \beta ([\mathcal{TB}]  +    [\mathcal{NB}]) } +   e^{  \beta (- [\mathcal{TB}]  +    [\mathcal{NB}]) }  \right)   \cr  \cr
Z^-_P={\rm tr}  \left(P e^{- \beta H^{-, \zeta}} \right) & \approx  &    e^{- \beta E_{0} }     2\sinh([\mathcal{TB}]\beta)\, e^{[\mathcal{NB}]\beta} \nonumber\\
&=& 
  e^{- \beta E_{0 }}   \left(  e^{  \beta ([\mathcal{TB}]  +    [\mathcal{NB}]) }  -    e^{  \beta (- [\mathcal{TB}]  +    [\mathcal{NB}]) }  \right)  
\end{eqnarray}
Therefore, the $[\mathcal{TB}]$ contribution to parity even and parity odd states has an over-all sign difference. As a result, we can write the leading non-perturbative contribution at energy level $N$ as 
\begin{align}
E_{N}^{\rm NP} &= -( e^{i \pi N}  [\mathcal{TB}]  +    [\mathcal{NB}])\cr
& = -( e^{i \pi N}   + e^{i \pi m}    )  [\mathcal{TB}] \cr
 &= -( e^{i \pi N}   + e^{i \pi m}    )   K_{N, m} e^{-2S_I},  
 \label{eq:main}
  \end{align}
 where $K_{N, m}  >0$. Therefore, the contribution of bion configurations to an energy level can be positive, negative or zero, depending on the even/odd parity of $N$ and $m$.   This result has the following implications: 
 \begin{itemize}
 \item {$m=$ odd, $N=2 \nu_{\rm outer}$= even (S): the NP contribution vanishes for such a symmetric state. These states are exactly solvable.}
 \item {$m=$ odd, $N=2 \nu_{\rm outer}+1$= odd (AS): the NP shift is  $E_{N}^{\rm NP} = 2K_N e^{-2S_I} > 0$.   The AS partner is lifted up non-perturbatively.  }
 \item {$m=$ even, $N=2 \nu_{\rm outer}$= even (S): the NP shift is $E_{N}^{\rm NP} = - 2K_N e^{-2S_I} < 0$.   The symmetric state is pushed down. These states are not exactly solvable. }
 \item {$m=$ even, $N=2 \nu_{\rm outer}+1$= odd (AS): the NP contribution is  $E_{N}^{\rm NP} = 0$.   These states are exactly solvable.}
 \end{itemize}
 This bion analysis explains the spectral structure in Section \ref{sec:qes-}, which was found using the BenderWu package.
Our construction provides a semiclassical path integral explanation of the remarkable phenomena observed in quasi-exactly-solvable  (QES) \cite{Turbiner:1987nw,Kozcaz:2016wvy}  systems.  In QES systems, for the $m=$odd (even)   case, a  number $\left\lfloor  \frac{m+1}{2} \right\rfloor  $ of symmetric (anti-symmetric)  states are exactly solvable and free of non-perturbative contributions. It was a puzzle why these states do not receive non-perturbative contributions despite the existence of obvious non-perturbative solutions. The above bion analysis, culminating in the expression (\ref{eq:main}),  resolves this puzzle. First, apart from the obvious configurations (the topological bions), there are also neutral bions. Their net effect  involves a subtle interference effect between the  hidden topological angle $\theta_{\rm HTA}$ and  the  discrete theta angle $\theta_{\rm disc.}$. 
When this expression vanishes in the path integral, the corresponding state  is exactly solvable in the Hamiltonian formulation.

\section{Conclusions}
\label{sec:conclusions}
We have shown that several new non-perturbative effects arise in the triple-well system that have no analogue in the familiar symmetric double-well system. The main results are the following.

\begin{itemize}
\item{In potentials with harmonic classically-degenerate minima not related by a symmetry, despite
the fact that  instantons are exact solutions,
they do not contribute to the energy spectrum at leading order in
semi-classics. The fluctuation prefactor of the instanton
amplitude vanishes if the  frequencies in two consecutive
well are not equal. }

\item{Quite generally the leading order semiclassical configurations contributing to the 
spectrum of such
systems are bion configurations. These are dominant configurations
that live on the Lefschetz thimble of critical points at infinity.}

\item{Even though the inner and outer wells have different shapes and
different curvatures, the perturbative expansions for states localized
in the inner and outer wells are related by an exact mapping; in the bosonic system 
(\ref{eq:pt-relation}) as well as in the 
supersymmetric  (\ref{eq:susy-pt-relation}) and quasi-exactly-solvable systems (\ref{eq:zeta_in_out}). This fact has a simple explanation in
terms of the all-orders WKB approach to perturbation theory. }

\item{In the SUSY and QES systems, there is an intriguing pattern of
interference between the neutral bions and topological bions, which arises from
the interplay of the hidden topological angle and the discrete theta angle.
Whenever the non-perturbative effects cancel precisely, the corresponding state in
the Hilbert space is exactly solvable. }

\item{ The bion analysis resolves an old puzzle concerning quasi-exactly-solvable systems. 
Despite the presence of obvious non-perturbative configurations which would contribute to the spectrum, the
spectrum turns out to be algebraic, and does not include non-perturbative factors. This is the result of interference between different bions, and the analysis shows that there also exist complex configurations, and there are exact non-perturbative cancellations
among them. This is a clear demonstration of resurgent structure in the SUSY and QES systems. }

\item{Semiclassical analysis based on the hidden topological angle $\theta_{\rm HTA}$
predicts that the character of the perturbative expansion
has an alternating pattern of convergent/divergent states, and then
changes from convergent to divergent after a certain energy level.
These predictions have been confirmed by a large-order analysis of the associated perturbative
expansions using the  Bender-Wu Mathematica package.\cite{Sulejmanpasic:2016fwr}}

\end{itemize}

Our bion construction shows that the semi-classical analysis for general quantum
potentials is far more intricate than the paradigmatic
textbook examples of the double-well potential and the periodic potential.
Perhaps, the most interesting lessons concern the important roles played by 
complex configurations and the remarkable interference patterns induced by
the hidden topological angle and the discrete theta angle. 
We believe that there are many other further  phenomena
waiting to be explored. In particular, it would be interesting to
investigate the appearance of the hidden topological angle, and the
interference between saddles, in the exact WKB formulation.

\section*{Acknowledgments}
This material is based upon work supported by the U.S. Department of Energy, Office of Science, Division of High Energy Physics under Award Number DE-SC0010339 (GD), and by the U.S. Department of Energy, Office of Science, Division of Nuclear Physics under Award DE-SC0013036 (MU). TS is funded by the Royal Society University Research Fellowship. We thank Yuya Tanizaki for useful discussions and comments.

\end{document}